# Roadmap for Rare-earth Quantum Computing

## Authors

Adam Kinos[1], David Hunger[2], Roman Kolesov[3], Klaus Mølmer[4], Hugues de Riedmatten[5,6], Philippe Goldner[7], Alexandre Tallaire[7], Loic Morvan[8], Perrine Berger[8], Sacha Welinski[8], Khaled Karrai[9], Lars Rippe[1], Stefan Kröll[1], and Andreas Walther[1*]

Affiliations:

1. *Division of Atomic physics, Lund University, P.O. Box 118, 221 00, Lund, Sweden*
2. *Karlsruhe Institute of Technology, Physikalisches Institut, Institute for Quantum Materials and Technologies, Wolfgang-Gaede Str. 1, 76131 Karlsruhe, Germany*
3. *3rd Institute of Physics, University of Stuttgart, Pfaffenwaldring 57, 70569 Stuttgart, Germany*
4. *Center for Complex Quantum Systems, Department of Physics and Astronomy, Aarhus University, DK 8000 Aarhus C, Denmark*
5. *ICFO-Institut de Ciencies Fotoniques, The Barcelona Institute of Technology, Mediterranean Technology Park, 08860 Castelldefels (Barcelona), Spain*
6. *ICREA-Institució Catalana de Recerca i Estudis Avançats, 08015 Barcelona, Spain*
7. *Chimie ParisTech, PSL University, CNRS, Institut de Recherche de Chimie Paris, 75005 Paris, France*
8. *Thales Research and Technology, 1 Avenue Augustin Fresnel, 91767 Palaiseau, France*
9. *Attocube Systems AG, Eglfinger Weg 2, 85540 Haar, Germany*

## Introduction and purpose

Several platforms are being considered as hardware for quantum technologies. For quantum computing (QC), superconducting qubits and artificially trapped ions are among the leading platforms, but many others also show promise, e.g. photons, cold atoms, defect centers including Rare-Earth (RE) ions. So far, results are limited to the regime of noisy intermediate scale qubits (NISQ), with a small number of qubits and a limited connectivity. In other words, we are still at the beginning of the technological development for QC. It is likely that future quantum technology hardware will utilize several existing platforms in different ways, taking advantage of their individual strengths. For these reasons, it currently makes sense to invest resources broadly and explore the full range of promising routes to quantum technology.

Rare-earth ions in solids constitute one of the most versatile platforms for future quantum technology. The central advantage is that the rare-earth ions are particularly resilient against some of the fluctuations from the environment, which enables them to demonstrate good coherence properties, even while they are confined in strong natural traps inside a solid-state matrix. This confinement allows very high qubit densities and correspondingly strong ion-ion couplings, i.e. an ideal situation for quantum information. In addition, although their fluorescence is generally weak, cavity integration can enhance the emission greatly and enable very good connections to photonic circuits, including at the telecom wavelengths, making them promising systems for long-term scalability.

---

[*] Corresponding author: andreas.walther@fysik.lth.se

RE systems have demonstrated world leading results in, e.g., quantum memories but have so far not demonstrated equally impressive results in QC. We believe that an important reason for this is that the resources spent on RE systems in the past decade are far below those spent on the currently leading platforms. There are, however, many reasons – from fundamental physical properties to technological accessibility – why QC with RE systems may provide a promising route for scalable quantum technologies. Thus, we argue that it will be worthwhile to spend the resources needed to drive the RE approach towards being a more mature technology.

The primary aim of this roadmap is to provide a complete picture of what components an RE quantum computer would consist of, to describe the details of all parts required to achieve a scalable system, and to discuss the most promising paths to reach it.

The first section provides an overview of the complete system and aims at providing an executive summary of the most important conclusions to readers outside of the rare-earth field of research. Sections 2-7 then provide in-depth discussions of each of the components and descriptions of the main mechanisms applied by the RE route to QC, its current status, and what can be expected from a near-future system based on realistic technological assumptions.

This document is the result of a community effort arising from the European Flagship project SQUARE[†] (Scalable QUAntum computing nodes with Rare-Earth ions), and the hope is that it will clarify the RE route to QC and act as a reference to compare with roadmaps of currently leading platforms as well as future advances of REQC.

---

[†] See http://square.phi.kit.edu/

# Contents

# 1 Overview and executive summary

This section gives an overview of the full scheme, and a summary of the most important results from the detailed sections below.

## 1.1 Why rare-earth ions? (pros and cons of RE versus other systems)

Solid state systems doped with rare-earth ions (REI's) were first suggested in the early 2000's for quantum information applications such as quantum memories (Nilsson & Kröll, 2005) and quantum computing (Ohlsson, Mohan, & Kröll, 2002). Since then, the RE parts of the quantum information field has been growing steadily, and now includes a large variety of quantum applications. In addition to quantum information processing and quantum memories, RE's are also promising candidates for photon/atom and microwave/atom interfaces. One of the most advanced goals of quantum technology, the quantum computer, is likely to consist of many of these different quantum components working together. Thus, the versatility of RE systems allowing them to act as any or all of the components, is a particular strength. Below, we highlight the main advantages of RE-based systems for quantum technologies:

- Best coherence properties of solid-state systems, combining a long optical $T_2$ of several ms (Könz, et al., 2003) with a long spin $T_2$, reaching up to 6 hours using strong magnetic fields (Zhong, et al., 2015).
- Ability to select single ions well below the focusing limit, by spectral addressing.
- Large number of spectral channels (unique qubit channels) due to a controllably broad inhomogeneous line.
- Close ion spacing allows strong qubit-qubit interactions via the dipole blockade mechanism
- More than nearest-neighbor interactions (estimated ~10 connections per qubit with current random materials)
- Variety of RE species at different wavelengths and with different spin properties, allowing varying interaction strengths as well as level structures.
- Crystals allow a variety of ways for integration into photonic and on-chip structures for production scalability.
- Transitions available at the main telecom wavelengths via Erbium (1.5 µm).
- Transitions available to allow coupling to microwave photons.
- Ensembles of millions of ions available for enhanced interaction with single photons, e.g. for quantum memories.
- Pseudo parallelism for all qubit operations, since swap operations and gate durations are compared to the optical $T_2$ (ms) during which other, idle, qubits barely decay due to much longer idle state hyperfine $T_2$ (seconds to hours).

While the advantages are clearly numerous, some come with caveats. For example, a long optical $T_2$ is fundamentally linked to a weak interaction with light. For quantum memories this can be easily circumvented using ensembles, but for a scalable quantum computer, reading out single ions would be required. The weak oscillator strength can be overcome with enhancing cavities, as is explained throughout this document.

## 1.2 The components of a scalable RE quantum computer

The are many options for all components of a rare-earth quantum computer, and they will be discussed in more detail later. However, here we would first like to summarize the most important findings, and give one concrete example of how a complete RE quantum computer could look like, using only components that can work together at the same time, and under realistic assumptions. The main findings are also summarized in Figure 1. This does not mean that a future REQC has to be built with these parts; indeed we expect that many of them could be improved or replaced by future work, but it provides a starting point to which other complete suggestions could be referenced.

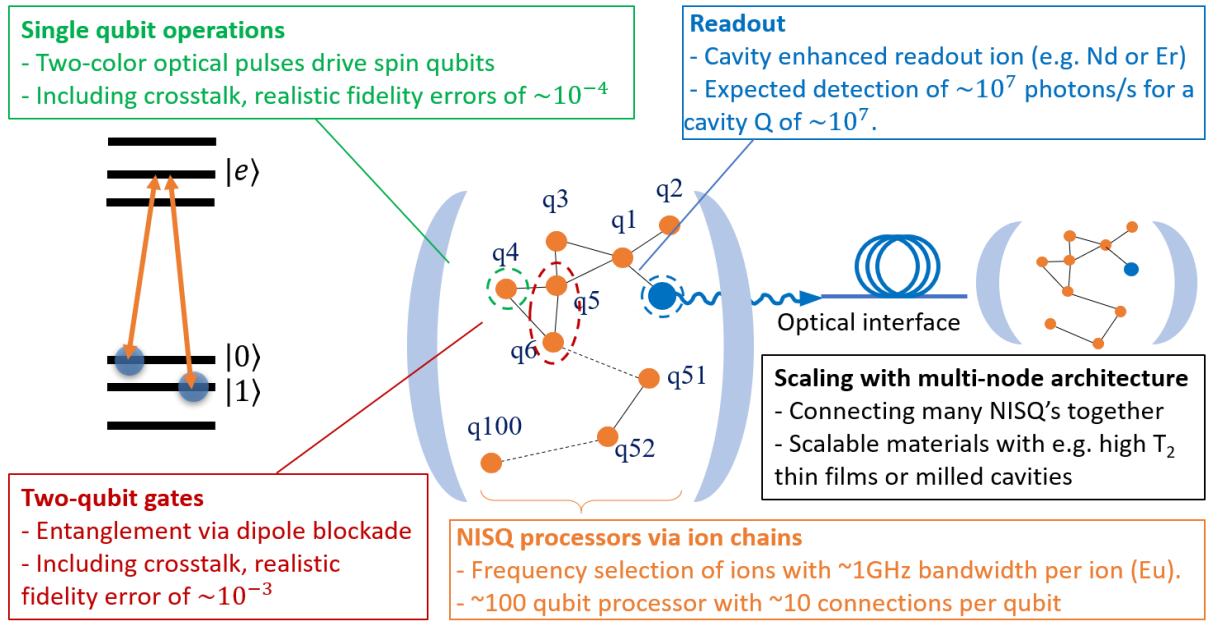

Figure 1. Overview of the components of a rare-earth quantum computer. Each component is described in detail in its own section. The numbers are estimations based on current technology and understanding, but it should be noted that many paths remain for optimizing them further.

**Overview** – The most promising REQC is based on using single RE ions, where one RE species is used as a long-lived qubit and another species can have its emission enhanced by a micro-cavity and used as readout. The dipole blockade mechanism can be used both to connect two qubits as well as connecting the qubits to readout, as described in more detail later. There are many possible combinations of RE species that could fulfill these roles, and in addition there are many possible host materials for them. Currently, this document will only focus on a few main combinations, which appears to have the best properties at this time. Although, it remains a future task to make more complete map of combinations, potential finding even better ones.

**Materials** – One material that could function in all roles is a $Y_2SiO_5$ crystal as host doped with $Eu^{3+}$ ions as qubits at a few percent level. In this crystal there would also be trace amounts of other RE ions at <1 ppm, and of those either Nd or Er could be used as a readout ion species. For future scalable production, such a material could be produced as a thin film, e.g. grown or etched down. Furthermore, ions could be placed deterministically by implantation. However, improvements of $T_2$ need to be made for the new solutions before they can be considered for a quantum computer supported by error correction. Currently, experiments show that a polished down bulk crystal could meet the requirements of good optical $T_2$ and readout (see discussion in Sec. 6.2). Therefore, this is the main concept described in this brief summary, although many material options exist for future improvements.

**Readout** – The emission from the readout ion would be Purcell enhanced by a high Q, low mode volume cavity to reach emission rates of the order of $10^7$ photons/s. This would translate into readout fidelities >99% (qubit state discrimination) in readout times of 10-100 μs. Readout ions without nuclear spin, i.e. a pure two-level system, can be chosen to simplify the cycling. In principle, different cavity types could reach high emission rates, and the type could be chosen due to other characteristics, such as the ability to scan to multiple spatial spots (fiber-cavities) or the ability for ns tunability (WGM-cavities) and integration on a chip. The latter can be extremely advantageous for high fidelity readout since both the single emitter and the superconducting detector (produced on the same chip) can be linked by an on-chip waveguide, therefore maximizing the photon collection efficiency.

**Single qubit gates** – Qubit states are coded as the nuclear spin of the ions in the material, and gate operations are performed via an optically excited state. This allows frequency addressing of qubits that sit a few nm's apart. The gate bandwidth needs to be significantly below the separation in frequency to other levels, both internal within the ions and external to other ions. Including these crosstalks, gate bandwidths below 10 MHz should allow single qubit gates with fidelity errors of the order of $10^{-4}$ with current schemes. Many options still remain to be explored for improving the fidelity beyond this value, especially those making more optimal use of the additional levels of the RE ions.

**Two-qubit gates** – Entanglement generation can be performed either via the electrical or the magnetic dipole-dipole interactions (dipole blockade); both can be of similar strength. Including crosstalk and restricting the bandwidth in a similar manner as for single qubit gates, two-qubit gates can be performed with fidelity errors of the order of $10^{-3}$, with current schemes and technology. This error is higher than for single qubit gates, since each two-qubit gate consists of several single-ion operations. In similarity with single qubit gates, the two-qubit gate fidelity could be significantly improved by novel gate protocols. In addition, future improvements to the micro-cavity technology could also allow new entangling mechanisms by means of the cavity mode. This could increase both the gate fidelity and the qubit connectivity by allowing more long-range interactions.

**Single node NISQ** – Using the dipole-blockade mechanism to sequentially find which ions can control the readout ion, and which in turn can control them etc., allows a set of connected ions to be found that can act like a NISQ processor node (Preskill, 2018). With a few percent doping concentration, and using a 100 GHz addressable inhomogeneous profile, a node of the order of 50-100 qubits can be included in such a near-term processor. The average connectivity per qubit will be about 10, though some ions will be more highly connected and some less. In such a node, all qubits could be uniquely addressed at a fast speed and operated with the fidelities given above. Future improvements to gate protocols, could make use of all qubit ion levels, which would allow a greater gate bandwidth not limited by the level splitting. In turn, this could allow the use of different qubit ions, such as Pr, which has smaller level spacing. This could significantly increase the number of qubits in the NISQ node, by decreasing the required frequency spread per qubit.

**Connecting many nodes** – One node with a readout ion in one cavity can be connected to the readout ion in another similar cavity, containing a similar NISQ qubit node, by means of detecting interfering photons. This can be performed in a heralded way to keep the entangling fidelity high. This step is similar to suggested ways of scaling for many other quantum computing schemes, such as artificially trapped ions or other defect centers, allowing ideas to be shared. RE ions trapped in nanomaterials may be particularly well suited to scaling by photonic means, owing to very high addressable qubit densities allowing miniaturization, and good photon connections, including at the telecom wavelengths.

## 1.3 Brief comparisons to other systems

Here we try to compare REQC to the two leading architectures of QC, namely ion traps and superconducting qubits. Such comparisons are inherently very uncertain, since research is always in a state of change, with new strengths discovered and weaknesses overcome. Nevertheless, the comparisons can help us learn how to view the platforms, and help us critically evaluate where they stand, and where they can improve. Although this is not the main purpose of this document, so it is kept very brief.

### 1.3.1 Brief comparison to ion traps

The rare-earth ions are what we call naturally trapped ions. They are trapped by the bonding potentials of the solid matrix, which corresponds to very high trapping strengths allowing the RE ions to sit only nanometers apart, and still be addressed through their spectral channel. This can be compared to ions trapped artificially by EM fields, which have trapping strengths that allow for micrometer separations. In the former case this allows the dipole-dipole interaction to be used which can give rise to strong interactions between more than just the nearest neighbor ions. For most schemes based on artificially trapped ions, qubit-qubit interactions using the motional states have to be used, which are difficult to scale within a single trap, though it should be noted that trapping Rydberg ions would also allow the dipole mechanism to be used (Mokhberi, Hennrich, & Schmidt-Kaler, 2020). One of the primary advantages of artificially trapped ions is that they are surrounded mostly by vacuum, which leads to a low level of perturbations. In contrast, RE ions reside inside a solid matrix where perturbations can sit closer, e.g. a nearby spin from the host. The close perturbations are mitigated by the famous property of RE ions that their active 4f levels are shielded by filled outer shells. A second aspect is efficient light collection. While trapped ions have strong readout transitions, it is challenging to integrate them with photonic structures due to perturbing electrostatic charges on dielectric surfaces. In contrast, RE ions have weak transitions which need strong cavity enhancement. With recent technological advances, integration of RE systems with micro- and nanophotonic cavities has been successfully demonstrated (Raha, et al., 2020; Dibos, Raha, Phenicie, & Thompson, 2018; Casabone, et al., 2020; Kindem, et al., 2020), with good prospects for the realization of scalable devices. With this, high readout fidelity now appears feasible, and the main strengths of the RE systems can shine through.

The most pursued route to QC scalability for artificially trapped ions is based on shuttling (Kaushal, et al., 2020), although there are several variants of this, including those with gates based on microwaves (Lekitsch, o.a., 2017). Since different schemes have slightly different pros and cons, it is hard to make valid comparing statements. However, generalizing a bit, the current schemes and results suggest that artificially trapped ions would be able to reach higher individual gate fidelities, but rare-earth systems would have more diverse and slightly stronger ways of connecting multiple qubits and computing nodes, and have a higher potential for miniaturization with compact cavities, crystals, and integrated photonics.

### 1.3.2 Brief comparison to superconducting qubits

For RE ions, the close spacing enables a very high qubit density per volume, where each qubit could require only about 5 $nm^3$. This can be compared to superconducting qubits, which struggles to go significantly below a mm in size per qubit. In addition, RE ions are addressed spectrally with light whereas superconducting qubits require wiring, some of it at very cold temperature (tens of mK), making it a serious logistical challenge to scale. In the face of the grand QC challenge, i.e. a technologically relevant algorithm that requires 100's of millions of qubits (Fowler, Mariantoni, Martinis, & Cleland, 2012), the size and addressability difference may prove important. To make a more sweeping comparison one would have to generalize the various schemes, but then it would currently seem that some advantages for superconducting qubits would include: more straightforward to add more qubits (at least early on), they are closer to conventional silicon technology, and they have a head start. However, RE system may have smaller footprints, efficient and low-noise optical addressing and readout, relaxed cooling requirements (1-4 K), and more diverse and stronger ways of connecting computing nodes, especially by optical means and by integrated photonics.

# Detailed description of all components

## 2 Cavity enhancement for efficient single ion readout

RE ions are promising for quantum information processing in large part due to their long life- and coherence-times. As a consequence, this gives them a weak emission rate for individual ions. For applications like quantum memories, ensembles can be used favorably to counter this by strong collective interaction using the high ion density of the RE system, but scalable QC relies on single ion operations. For single ions, the intrinsically weak light-matter interaction can be drastically increased by using a variety of techniques centered around the enhancement provided by wavelength-sized high quality factor cavities.

Due to the Purcell effect, an ion resonantly coupled to a cavity with a high quality factor $Q$ and a small mode volume $V$ will emit at an accelerated rate $\gamma = (F_P + 1)\gamma_0$, where $\gamma_0$ is the free-space emission rate and $F_P = \zeta \frac{3}{4\pi^2} \frac{Q}{V/\lambda^3}$ is the *effective Purcell factor*. Here we include the branching ratio $\zeta$ of the relevant transition (ranging between ~1% for $Eu^{3+}$:$Y_2O_3$ and 45% for $Tm^{3+}$:$LiNbO_3$), such that $F_P$ corresponds to a lifetime reduction factor. In contrast, $F_P/\zeta$ describes the emission enhancement for a particular transition and can be referred to as the *ideal Purcell factor*. Purcell enhancement increases the emission rate as well as the collection efficiency, since a fraction $\beta = F_P/(F_P + 1)$ is emitted into the cavity. It is important to design cavities which efficiently couple out the light, ideally into a single spatial mode for detection or further use within a network. A high outcoupling efficiency $\eta = \kappa_x/\kappa$ requires the cavity outcoupling rate $\kappa_x$ to be as large as possible compared to the overall cavity decay rate $\kappa$. In addition, $\kappa \gg \gamma_0$ is desired, and essentially always fulfilled for the small values of $\gamma_0$ for RE ions.

Purcell enhancement can furthermore restore photon indistinguishability which is important e.g. for entanglement generation between remote qubits. The intrinsic degree of indistinguishability $\epsilon^0 = T_2/2T_1$ is given by the ratio of the total dephasing rate $1/T_2 = 1/2T_1 + 1/T_2^*$ and the excited state lifetime $T_1 = 1/\gamma_0$. For most RE-doped materials and in particular nanomaterials, $T_1 \gg T_2$, and $F_P > T_2/T_1$ is required to increase the indistinguishability to $\epsilon^c = T_2/2T_1^c$ by reducing the lifetime in the cavity $T_1^c = 1/(F_P + 1)\gamma_0$.

Finally, the Purcell factor is proportional to the cooperativity $C = \epsilon^0 F_P/2$, which quantifies e.g. the extinction contrast of a single ion coupled to the cavity. For many quantum network schemes based on single photon absorption or reflection for quantum state transfer or entanglement generation, $C \gg 1$ is required to achieve high fidelities.

### 2.1 Readout-ion scheme

The best performance for qubit readout is expected to be achieved when possibly several qubit ions are coupling to a separate readout ion, and the latter is Purcell enhanced via the cavity (see also Secs. 3, 6.3). With this, readout fidelities can be maximized and excited states of the qubit ions remain unaffected, which is desirable for qubit gate operations. Readout ion transitions with a short excited state lifetime and large branching ratio are beneficial (McAuslan, Longdell, & Sellars, 2009), however, the transition linewidth is required to remain narrow enough to enable high-fidelity gates with the qubit ions. A long transition wavelength - possibly telecom wavelength to allow for low loss long distance interfacing – is advantageous. Long wavelengths are furthermore beneficial for cavity fabrication because scattering and absorption loss decrease with wavelength, the former with $\lambda^{-2}$. Also fabrication tolerances are relaxed, such that the highest quality factors and thus Purcell factors can be achieved. A natural choice is to use Erbium as a readout ion. However, the combination of a long optical $T_1$ (~10ms), a moderate branching ratio ($\zeta$~0.2), and the need for lowest ion concentrations to avoid magnetic ion-ion interactions make Erbium also a challenging candidate. Neodymium has a strong transition, but shares the drawbacks which arise from the presence of an electron magnetic moment. Recent results with Ytterbium are promising (Ortu, et al., 2018; Kindem, et al., 2020) due to insensitive clock transitions at zero magnetic field. As a qubit ion, Europium stands out due to its long optical and spin coherence time and level structure, which makes it particularly suitable for a scalable quantum computing scheme as detailed in the later chapters. Combining Erbium as readout and Europium as qubit is thus a promising approach (Kimiaee Asadi, et al., 2018), while using Europium for both the qubit and the readout represents a simpler alternative. Due to the multifaceted situation, we thus consider here cavity enhancement of all relevant ion species.

In the following sections, the main cavity architectures are described, listing the pros and cons of the different methods, while commenting on their scalability.

## 2.2 Open-access Fabry-Perot microcavities

Open access microcavities consist of two separate highly reflective mirrors that allow one to introduce various material types such as nanocrystals, thin films, or membranes into the cavity mode in a controlled manner. Different cavity geometries have been used, including cavities combined from a single-mode fiber and a planar substrate carrying the sample, micromachined planar substrates, and fully fiber-based systems. The concave mirror profiles that define the lateral cavity mode can be produced by $CO_2$ laser machining (Hunger, o.a., 2010), focused ion beam milling (Trichet, al., & Smith, 2015), or plasma etching (Wachter, al., & Trupke, 2019). The techniques produce near-spherical geometries with radii of curvature down to a few micrometers (Kelkar & et al., 2015; Trichet, al., & Smith, 2015), leading to cavity mode volumes as small as a few wavelengths cubed (Benedikter, et al., 2017; Najer, al., & Warburton, 2019; Kelkar & et al., 2015; Trichet, al., & Smith, 2015). The quality factors are not a fixed quantity for such cavities but depend on the mirror separation. Recent state-of-the art values in experiments with RE ions in the cavity are: a mode volume $V = 9(\lambda/n)^3$ and a quality factor of $Q = 1.2 \times 10^5$ at the shortest length at a wavelength of 580nm with $Eu^{3+}:Y_2O_3$ nanocrystals (Casabone, o.a., 2018), leading to a Purcell factor $F_P = \zeta \times 1000$. Improved parameters have been achieved with $Er^{3+}:Y_2O_3$ nanocrystals in the telecom band ($V = 4.6(\lambda/n)^3$, $Q = 9 \times 10^4$, $F_P = \zeta \times 1500$) (Casabone, o.a., 2020), and current record values are reported for a different material system, epitaxial quantum dots at 980nm ($V = 1.4\lambda^3$, $Q = 1.5 \times 10^6$, $F_P = \zeta \times 8 \cdot 10^5$) (Najer, al., & Warburton, 2019).

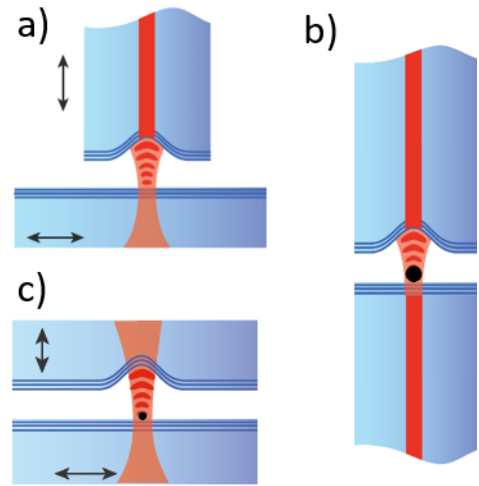

Figure 2. a) Fiber-based scanning Fabry-Perot microcavity based on a single-mode fiber and a planar substrate. b) Integrated version based on two single-mode fibers where only the cavity length remains tunable. c) Open-access scanning microcavity based on planar substrates.

Limiting factors are scattering and absorption losses of the mirrors, which strongly increase towards shorter wavelengths. While at telecom wavelength, $Q$~$10^7$ can be anticipated, cavities at visible wavelengths as required e.g. for $Eu^{3+}$ may at best approach $Q$~$10^6$. Also, scattering due to nanocrystals or rough surfaces from thin films need to be accounted for. However, restricting the crystal size to ~60 nm (Liu, et al., 2020) as well as polishing techniques (Merkel, Ulanowski, & Reiserer, 2020) have already reduced these losses to an unproblematic level.

Assuming optimal values for $Q, V$, one can expect Purcell factors up to $F_P \sim \zeta \times 10^5$ in the visible and $F_P \sim \zeta \times 10^6$ in the near infrared. For the three example species $Eu^{3+}$, $Nd^{3+}$ and $Er^{3+}$, the expected enhancement would approximately be $F_P = 10^3$, $10^4$ and $10^5$, respectively. Currently available nanocrystals approach $\epsilon^0 = 5 \times 10^{-3}$ (Eu) and $5 \times 10^{-5}$ (Er) at 3K, such that in both cases, high indistinguishability $\epsilon^0 \approx 1$ and cooperativity are within reach. Material improvements or lower temperatures can be expected to offer further increase of $\epsilon^0$ by increasing $T_2^*$. In an alternative approach, using a polished membrane of a high-purity bulk YSO crystal, lifetime limited optical coherence of $Er^{3+}$ with $\epsilon^c \approx 1$ has already been achieved (Merkel, Ulanowski, & Reiserer, 2020). In this experiment, using a cavity with planar substrates and $Q \sim 10^7$, a Purcell factor $F_P = 60$ was demonstrated for ensembles, such that $C \gg 1$ is expected for single ions.

Current challenges of the fiber cavity approach are polarization control and mode matching for efficient fiber-based outcoupling. A general difficulty is the control of resonance conditions under cryogenic conditions in the presence of mechanical noise, but compact custom designs have recently allowed operation with sufficient stability (Merkel, Ulanowski, & Reiserer, 2020; Casabone, et al., 2020).

Specific advantages of the open-access design are the possibility to study different samples with one and the same cavity, to dynamically control the resonance frequency at high bandwidth e.g. to address several ions at different frequencies with possibly up to microsecond switching speed (Casabone, o.a., 2020), to achieve frequency selectivity for specific transitions e.g. to induce cycling transitions, and efficient free-space outcoupling (>90% is possible). This cavity type also allows to use the smallest nanomaterials, which is a crucial advantage when targeting highly doped samples that we envision here, since small scales avoid spectral crowding in the available frequency channels, and the probability to find strongly interacting ions is large. Scalability can be envisioned by fully fiber-based devices which can be compact and require minimal control resources.

### 2.3 LiNbO$_3$ WGM cavity

An approach based on single crystal thin films of ferroelectric materials activated with REIs offers several advantages such as a small footprint of the chip surface (tens of µm$^2$ or smaller), scalability of production (thousands of devices can be produced on the same chip), ability of targeted activation with REIs by focused or masked ion implantation, and reasonable tunability by fast electric fields (0.1-0.2% of the carrier optical wavelength). The latter feature is particularly important since the ferroelectric cavity can be tuned in and out of resonance with the REIs at sub-nanosecond timescale therefore allowing for dynamic control of Purcell enhancement.

RE ions doped by implantation into thin film lithium niobate disk resonators has been investigated. The disk geometry is chosen for the simplicity of fabrication. Disk resonators have moderately low mode volume of 50 $(\lambda/n)^3$ and rather high loaded Q-factor of 2-3×10$^5$. Yb$^{3+}$ ions implanted into the resonators at low concentration have spectral properties similar to the ones in the crystals doped during growth (Kis, o.a., 2014) as figured out in the course of SQUARE by spectral holeburning (Zauerzapf, 2020). As of now, five-fold Purcell lifetime shortening of Yb$^{3+}$ emission was measured in moderate high quality resonators. The emission is directly coupled out of the resonator by a predefined monolithic waveguide placed next to the cavity and evanescently coupled to it. Cavity tuning is accomplished by a top electrode placed on top of the cavity with a thin film dielectric spacer. The tuning range of the TE-mode is defined by the r$_{31}$=8.6 pm/V electro-optic coefficient and was measured to be 240 GHz around 980 nm wavelength.

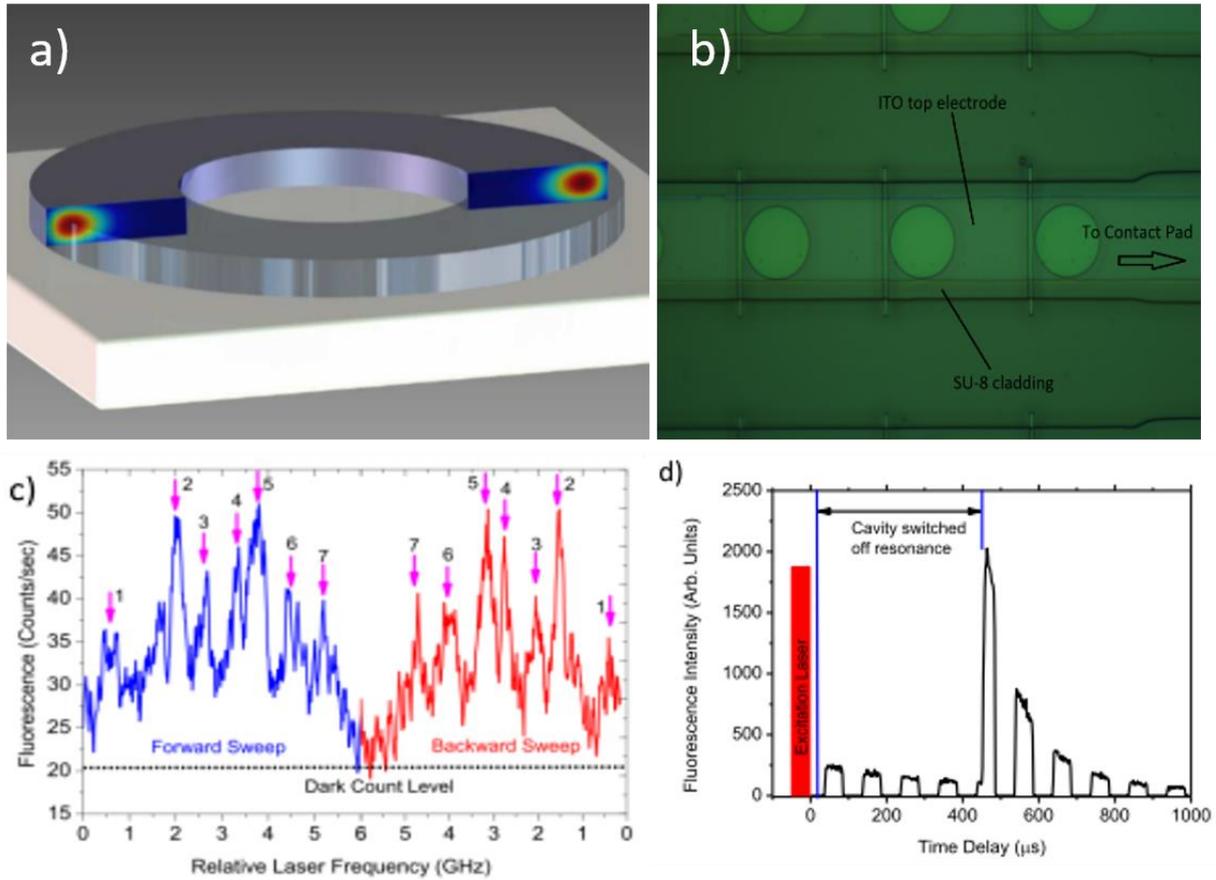

*Figure 3. a) Schematic drawing of a thin-film ring cavity with simulated electric field distribution. b) Fabricated $LiNbO_3$ disc cavities evanescently coupled to waveguides and capped with electrodes for tuning. c) $Yb^{3+}$ resonant fluorescence as the laser frequency is swept through the cavity resonance. The cavity is tuned to the shoulder of the inhomogeneous line of $Yb^{3+}$. The peaks corresponding to 1-2 ions are marked. The symmetry of the peaks for forward and backward sweeps can be clearly seen. d) Storage and retrieval of optical excitation as the cavity is tuned off and back on resonance with the excited $Yb^{3+}$ ions. The fluorescence is chopped by an optical chopper. That explains periodic dips in the signal. The fluorescence is retrieved 400μs after the cavity is tuned off resonance.*

Current Q-factors and mode volumes allow for the emission speed-up by a factor of 50-100 taking into account the emission branching ratio of the $Yb^{3+}$ ion. Having in mind 400 μs natural lifetime of $Yb^{3+}$ emission in $LiNbO_3$, the emission rate of $12-25\times10^3$ photons/s is expected. For critically coupled cavities, half of the photons (i.e. $6-12\times10^3$ s$^{-1}$) end up in the outcoupling waveguide and can be used to detect single $Yb^{3+}$ ions. Taking into account the current 30% extraction of photons into free space, 70% transmission of the objective lens, and 60% of the detector efficiency, one expects 750-1500 detected photons per second out of a single ion. Recently, small sub-ensembles (1-2 ions) of $Yb^{3+}$ ions were spectrally isolated in not so high-Q $LiNbO_3$ cavities at the University of Stuttgart. Furthermore, we obtained fast (a few microsecond) tuning of the Purcell enhancement allowing for excitation storage and it's on demand release. The data is being prepared for publication.

Disk resonators are not the only possible geometry for realization of Purcell enhancement of the emission of REIs. Photonic crystal cavities with Q-factors up to $10^7$ and mode volume on the order of 1.4 $(\lambda/n)^3$ were demonstrated in silicon-on-insulator films (Asano, Ochi, Takahashi, Kishimoto, & Noda, 2017). If similar cavities are realized in $LiNbO_3$ thin films, Purcell speed up of the fluorescence of $10^5$ is expected resulting in a 4 ns lifetime of $Yb^{3+}$ emission, giving a detection rate of $10^6$ photons/s with similar assumptions as above. Since the linewidth of the cavity mode is only 30 MHz, i.e. much smaller than the bandwidth of the enhanced $Yb^{3+}$ emission, a high degree of indistinguishability for single $Yb^{3+}$ can be achieved. At the same time, the coupling strength can be dynamically controlled allowing for excitation storage when the cavity is off resonance with the emitter and for release of the photon once tuned in resonance with it. Thus, deterministic single photon sources with on-demand emission can be realized thanks to long intrinsic lifetime of REIs and dynamically tunable high Purcell enhancement. This option paves the way to implement quantum technology with high complexity and high degree of integration.

$LiNbO_3$ is a material with very dense and heavy nuclear spin bath. This fact makes it quite unfavorable for quantum information processing. However, at least two ways around this problem are feasible. First, instead of using naturally abundant isotopes of $Yb^{3+}$, one can implant $^{171}Yb$ having ZEFOZ point in zero magnetic field due to hyperfine interaction between $^{171}Yb$ nucleus and its electron. This strategy is similar to the one exploited in (Kindem, o.a., 2020). Implantation of $^{171}Yb$ into $LiNbO_3$ thin films is currently under development in collaboration with University of Bochum. The second option is switching to a spin free ferroelectric host material, i.e. $BaTiO_3$. The latter approach gives more advantages in the future though requires extensive material study to make single crystalline thin films.

## 2.4 Focused-Ion-Beam milled cavities

Directly shaping REI-doped crystals with a focused-ion beam has proven to be a very successful approach to realize nanophotonic cavities with high Purcell factors. The geometry of a nanobeam with triangular cross section and periodic grooves to realize Bragg mirrors achieved wavelength-scale mode volumes ($V = 1(\lambda/n)^3$ and good quality factors ($Q = 1 \times 10^4$) at 980 nm for $Yb^{3+}$ (Kindem, o.a., 2020). Together, this leads to large Purcell factors up to $F_P = \zeta \times 350$. Numerical simulations predict higher values with up to $Q = 10^5$, and $V \sim 0.5(\lambda/n)^3$, leading to

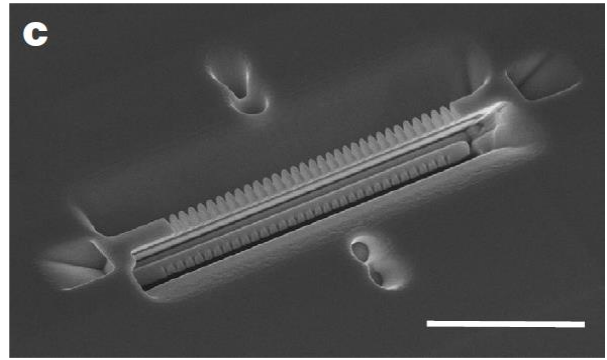

*Figure 4. Scanning electron micrograph of a photonic crystal cavity produced by FIB milling in $YVO_4$. Scalebar 10µm. Taken from (Kindem, o.a., 2020).*

Purcell Factors up to $F_P \sim \zeta \times 10^4$. This could be achieved by improved fabrication quality or by operating at telecom wavelengths and coupling to Er ions. However, at visible wavelength, scattering loss currently seems to be too large to achieve high Purcell factors. Furthermore, Ga incorporation and lattice damage may lead to fundamental limitations in the coherence properties of the ions, although to date, such influence has not been evidenced. In contrast, Fourier-limited photons of Purcell-enhanced single Nd ions was demonstrated recently (Zhong T., et al., 2018), and good optical ($\epsilon^c = 0.05$) and long spin coherence ($T_2^{DD} = 30 \; ms$) of single Yb ions (Kindem, o.a., 2020). Highly efficient outcoupling into a low-NA free-space mode can be achieved by 45° undercut output couplers, limited essentially by the quality of the collecting lens and the Fresnel reflection at the crystal – air interface.

## 2.5 Nanophotonic cavities with evanescent coupling to RE's

Shaping nanophotonic cavities from low-loss photonic materials such as silicon at telecom wavelength or SiN in the near infrared has achieved the highest Purcell factors (Dibos, Raha, Phenicie, & Thompson, 2018) (Raha, o.a., 2020). Record values of $V = 0.3(\lambda/n)^3$, $Q = 6 \times 10^4$ have led to $F_P = \zeta \times 4500$ and a corresponding lifetime reduction by a factor 700 was demonstrated with Er ions. To couple to the sample, bonding on high-quality bulk crystals is performed. While this reduces the electric field at the location of the ions since only the evanescent tail can be used, it allows one to optimize the fabrication independently of the sample material, and to make use of ultra-pure materials. Cavity fabrication can be scalable, and further improvements of the design and fabrication quality could lead to $Q \sim 10^7$ (Asano, Ochi, Takahashi, Kishimoto, & Noda, 2017) (Vasco, Gerace, Seibold, & Savona, 2020), and cascaded slotting to $V \sim 10^{-4} (\lambda/n)^3$ (Choi, Heuck, & Englund, 2017), such that even if these two extreme values can't be achieved simultaneously, $F_P \gg \zeta \times 10^6$ is expected to be achievable. Although it should be noted that, so far, no measurements of long $T_2$ have been made with this configuration.

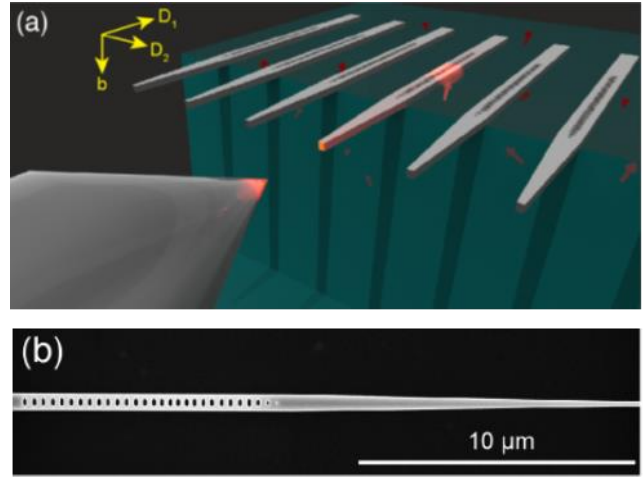

Figure 5. a) Schematic drawing of Si photonic crystal cavities evanescently coupled to Er ions in a YSO bulk crystal with adiabatic outcoupling to a tapered fiber. B) SEM image of a cavity. Taken from (Dibos, Raha, Phenicie, & Thompson, 2018).

Silicon cavities can be used only for RE emitters having wavelengths longer than 1.1 μm (the bandgap of Si), i.e. Erbium ions. However, other materials, such as diamond or silicon carbide membranes, gallium nitride (GaN) thin films, and a few other materials can be used to devise cavities for wavelengths in the visible and even in the near UV regions. Evanescent coupling in a cavity made of these materials would allow access for gate pulses targeting almost any REI, e.g. $Eu^{3+}$ or $Pr^{3+}$, i.e. the ones with the best coherence properties. When the material processing technology develops to as mature state as the one for silicon, the evanescent coupling scheme will look very attractive also in the visible domain.

## 2.6 Readout fidelity

To determine the effect of the readout step on a quantum computer, it is important to estimate both the readout fidelity and the duration, since the readout step should not be much worse than typical gate operations. What can be obtained naturally differs somewhat between the cavity technologies. For this analysis, we assume a cavity Q value of $10^7$, which would yield an enhanced lifetime of about 100 ns for either of the envisioned readout ions, Er or Nd, when placed in a Fabry-Perot cavity, as described above. In a typical setup, not all photons would reach the detector. However, due to the Purcell enhancement, almost all photons are emitted into the cavity mode. Furthermore, the coupling to the fiber can be made reasonably efficient (Hunger, o.a., 2010) and detector efficiencies for the considered wavelengths can be 80% for superconducting detectors. Therefore, it is reasonable to assume that at least 10% of all emitted photons can be detected, giving a detection rate of $10^6$ photons/s.

One way to achieve a state selective readout is to use cavity enhancement to ensure cyclicity during excitation and emission on the qubit ion directly. This was demonstrated in Er with 95% state discrimination fidelity using a readout duration of 20 ms (Raha, o.a., 2020). Similar results have also been demonstrated in Yb with a state discrimination fidelity of 95%, although this required post-selection that had a success rate of 64% (Kindem, o.a., 2020).

In order for the cavity enhancement not to cause decay on the qubit excited states however, the main scheme to achieve state selective readout is by having separate qubit and readout ion species. Then, one can detect the state of a qubit ion that is close to the readout ion, by exciting one of the spin qubit states to the exited state, which will shift the levels of the readout ion (or not if the other state was instead populated), i.e. using the dipole blockade mechanism. This means that state detection can be made by separating the photon counts for ion emission vs. not emitting (only background).

The optimal strategy for using the information from the detected photons, including the information about the timing between events, is using Bayesian statistics, and a protocol that describes how to apply this to our systems is given in (Debnath, Kiilerich, & Mølmer, 2020). In that paper, the decay rate is replaced by our detection rate (in most places), and to separate the states, the interaction shift needs to be significantly larger than 1/decay rate. In our case, it can be expected to reach interaction shifts of 10-100 MHz (see Sec. 3.2). Assuming a shift of about 5 times the decay rate, and a qubit lifetime of about $10^4$ times the decay rate (Eu lifetime ~1 ms), Ref (Debnath, Kiilerich, & Mølmer, 2020) suggests that the readout fidelity will saturate at about 95%, after about 10 times 1/detection rate, i.e. after about 10 μs. This saturation is caused by the finite qubit lifetime, since if this undergoes spontaneous decay during the detection, the state information is lost.

The problem with finite lifetime can be mitigated by using a buffer ion, which is described in detail in Ref (Walther, o.a., 2015). Here, one qubit ion is taken as a dedicated buffer to the readout ion, to which the (classical) state of any other qubit that should be readout can be transferred. Thus, any decay event on the buffer ion does not destroy the state information of the qubit. By cycling only a few times between the qubit and the buffer ion, the state discrimination fidelity can be improved by more than one order of magnitude to ~99.9%, at the cost of extending the duration of the detection to ~40 μs. This fidelity is now mainly limited by decay during the transfer pulse from the qubit to the buffer ion.

## 3 High fidelity qubit gates

The main reason why we believe that it should be possible to build a quantum computer is the ability of quantum error correction to mitigate the effects of errors (loss of fidelity) in quantum gates. There are several protocols for performing quantum error correction, which require different threshold values of the gate fidelity to work, add a different amount of overhead in terms of more qubits, and which in turn grant different error reduction gains. Although not necessarily the best in all situations, one of the standard schemes for quantum error correction is the Steane code (Steane, 1996). It is believed that the threshold for gaining any error reduction by applying the error correction, compared to a single physical qubit, is a fidelity better than 99.99% or conversely a fidelity error lower than $10^{-4}$. Some other protocols, like surface codes (Bravyi & Kitaev, 1998), can work with threshold errors up to $10^{-2}$, however at the expense of significantly higher number of overhead qubits (Fowler, Mariantoni, Martinis, & Cleland, 2012). As an example, using surface codes with error rates a factor of 10 below the threshold might require approximately 10,000 physical qubits per one logical qubits, in order to run a reasonable algorithm.

These fidelity thresholds provide an obvious target that a potential QC should aim for, as the first step in order to reach the regime where testing error correction and running algorithms is possible.

The fundamental limitation to the fidelity comes from the duration of the gate pulses in relation to the coherence time. However, in practice the fidelity may also be limited by other external mechanisms such as fluctuations in the gate pulses or the environment. In addition to improving fidelity, fast gates will also yield a short algorithm running time. Thus, it is important to achieve as fast gates as the available bandwidth allows. In this section, we go through the various gate schemes that are suitable to the RE systems and discuss their potential.

## 3.1 Single qubit gates
### 3.1.1 Protocols for single qubit gates

There are several different protocols that can be employed to perform gates in RE systems. Here, a few of them are mentioned and briefly described together with the main advantages and disadvantages.

*Dark state scheme*

The dark state scheme for performing gates on spin qubits mediated by optical transitions was suggested in Ref. (Roos & Mølmer, 2004), and demonstrated experimentally for RE ensembles (Pr) in Ref. (Rippe, Julsgaard, Walther, Ying, & Kröll, 2008), reaching a fidelity of about 92%. The scheme makes use of bichromatic pulses targeting a lambda type scheme, where the two ground state levels are the qubit states. The interference between the different paths of the multiple frequency components creates an effective dark state and a bright state among the qubit levels, where only the bright part of the wavefunction interacts with the light. Ref. (Roos & Mølmer, 2004) explains how a pair of perfect transfer pulses between the selected bright state and the excited state yields an arbitrary gate on the qubit levels. This protocol can be combined with pulse-shapes, like the chirped, complex hyperbolic secant shape (sechyp), on the light field components in order to achieve robustness against certain experimental variations such as power fluctuations and differences in optical resonance frequency. The latter is particularly important when used in ensembles, since they consist of inhomogeneously broadened selections of ions. In the original protocol, a single qubit gate using 4 bichromatic pulses, and a two-qubit gates using 12 such pulses were shown to yield high performance on qubits with a rather broad inhomogeneous broadening. However, several of these pulses were used to compensate for difference phase shifts acquired by the detuned ensemble, when the bright and dark states where excited at different times. In the single ion regime, where the resonance frequency may be determined with sufficient accuracy these extra pulses are not required. This brings the number of bichromatic pulses down to 2 for single qubit gates and 4 for two-qubit gates.

The robustness of this technique comes at a price of the long duration of the pulse sequences, which reduces the fidelity due to decay and dephasing.

*Simple cut pulse shapes and DRAG*

In the single ion regime, the robustness offered by the sechyp-shaped pulses may not be required. The simplest pulse shapes are square pulses, Gaussians or cut Gaussians. Square pulses make optimal use of the available power during a short duration, but they have broad spectral components which can excite other ions and cause direct errors or spectral diffusion. Gaussians are much smoother with suppressed power outside the frequency range of the targeted qubit transition, but they need to be fairly long not to suffer from truncation errors. A trade-off between the two are suitably cut Gaussians, that are displaced in height so that they are turned on and off in a continuous manner. These are quite efficient in terms of power use, and they still suppress higher frequency components, making them favorable in our system.

One error in RE systems is the internal crosstalk to other levels of the same qubit. For direct transition two-level qubits, such as using two of the levels of an-harmonic oscillators of e.g. superconducting qubits, it has been shown that simple cut Gaussians can be adapted by using frequency chirps, off-quadrature or both to negate the excitations to the unwanted levels, in a scheme known as Derivative

Removal by Adiabatic Gate (DRAG) (Motzoi, Gambetta, Rebentrost, & Wilhelm, 2009). However, for more indirect qubit systems, such as the lambda-type schemes of the RE ion spin qubits, there is no known way to deduce a simple compensation. A numerical method has been used at Lund University, in order to find chirps/detunings that provides some compensation, although the added benefit over the simple cut Gaussian is so far low, as is discussed in Sec. 3.1.2.

Including the main error sources, optical decoherence, internal and external crosstalk, a single qubit gate fidelity error of $2 \cdot 10^{-4}$ was achieved for DRAG pulses in full level simulations (Lund University, 2021). Here, $^{153}$Eu:YSO was considered, which we believe is the best candidate for a RE qubit. More details and a longer discussion on the error sources is given below in Sec. 3.1.2.

*Schemes based on numerical optimization (Optimal control theory or shortcut-to-adiabaticity)*
The schemes described above are all based on intuitive and semi-analytical reasoning for finding pulse shapes and sequences. The experience from other quantum computing platforms is that numerical search methods may improve significantly on the fidelity and duration of gates (Choi, et al., 2014) (Goerz, Motzoi, Whaley, & Koch, 2017) in particular as they may employ dynamics that explores - rather than meticulously avoids - intermediate population of auxiliary levels and transition path ways. There are a few different techniques to explore such processes, the semi-analytical shortcut-to-adiabaticity strategy and fully numerical optimal control theory.

For RE ions, optimal control theory has been tested for Pr ensemble qubits (Walther, et al., 2009). Although these preliminary tests did not improve the pulse fidelity over other pulse-shapes, more extensive tests have to be done, in particular aimed at the single ion regime. Pulse optimization based on shortcut-to-adiabaticity has also been tested in RE systems (Yan, et al., 2019), although only using more specialized initialization pulses rather than general purpose qubit gates. The fidelity was tested experimentally on Pr ensembles and a fidelity of about 98% was achieved, compared to about 92% for the uncorrected sechyp dark state gates, but we believe that the restriction to a specific initial pulse (sechyp in our case) is not the best starting point for numerical search for optimal gates.

*Direct driving by radio frequency (RF) fields*
Directly driving the spin levels by using RF fields may give a high fidelity even for single spins. Single qubit gate errors below $10^{-5}$ were experimentally demonstrated using microwave driven NV centers in diamond (Chou, Huang, & Goan, 2015). However, as we will see in the next section, scalability by means of dipole blockade and frequency addressing will no longer work in a straightforward manner, since the spin inhomogeneity is much smaller than the optical inhomogeneity, typically by ~six orders of magnitude, i.e. tens of kHz rather than tens of GHz. This means that only very few unique qubits could be addressed spectrally and in addition the pulses would have to be very long not to excite anything outside their targeted bandwidth, which would cause the operations to be very slow. Therefore, these types of gates could only be exploited as primary qubit gates together with a different scheme for scaling and multi-qubit addressing. Although it may also be possible to use RF driving as a compliment to other gates, e.g. accomplishing dynamic decoupling sequences broadly at many qubits in parallel.

### 3.1.2 Fidelity dependence on pulse and material properties
In the previous section a number of techniques for finding pulse shapes and sequences to realize qubit gates were presented. Regardless of which technique is used, there are some common physical mechanisms that set the limitations to the achievable fidelity. In this section, these limitations are listed and discussed. We will focus on the details on the primary candidate for qubits for a RE QC, which is site 1 of the $^{153}$Eu species, doped into $Y_2SiO_5$.

The main reasons are that it has a demonstrated long coherence time both in the optically excited state (Könz, et al., 2003) and in the spin levels (Arcangeli, Lovric, Tumino, Ferrier, & Goldner, 2014). In addition, it has a large separation between the optical levels, which allows a larger gate bandwidth to be used, translating into a shorter pulse with lower error from dephasing. Furthermore, site 1 offers a more balanced set of oscillator strengths between the transitions, which allows for minimizing the error sources discussed below:

*Gate duration compared to the $T_2$ of the optically excited states used to mediate the gate.*
The fidelity error of the gate scales exponentially with the gate duration, and can be approximated as $F_{err} \sim e^{-t_{dur}/T_{2,optical}}$. The hyperfine $T_2$ of course also needs to be sufficiently long, although since this affects the qubits at all times, not just during gates, the hyperfine $T_2$ in practice has to be much longer than the optical $T_2$. This can be accomplished either by aligning the magnetic fields to a ZEFOZ point, and/or by using dynamic decoupling sequences. The former has the disadvantage of splitting all levels which might limit the available bandwidth. Dynamic decoupling sequences however, could most likely be incorporated into idle times for qubits through proper compiler sequencing, and should enable the hyperfine $T_2$ to be sufficiently long (Souza, Álvarez, & Suter, 2011; Fraval, Sellars, & Longdell, 2005). For Eu:YSO, an optical coherence time of 1.5 ms without magnetic field has been achieved, which limits the maximal duration to a few microseconds, if this should not be the main limitation.

*Pulse bandwidth compared to internal ion levels (internal crosstalk)*
Depending on which species of RE is used, there are a different number of levels in the ground and excited state. For the primary qubit candidate, Eu, there are three hyperfine levels in each electronic state. Two of the ground state levels are used as qubit states, and one excited as mediator. However, when any light field is turned on to perform operations, there is a chance to have off-resonant interactions with other ions, which can cause either bit errors or phase errors. We refer to this as internal crosstalk, and this contributes to decoherence of the qubit gate. For Eu, the separation of the other levels is of the order to ~100 MHz, which limits the bandwidth of the pulses to the order of ~10 MHz if this should not be the main limitation.

*Pulse bandwidth compared to spectral channel of other ions (external crosstalk)*
The off resonant excitation of other ions will also contribute to decoherence. If an ion that is spatially close to the qubit ion gets off-resonantly excited, it will change the energy level of the qubit ion via the dipole-dipole interaction mechanism (same as is used for entanglement described later). This will cause dephasing during the gate duration. At an ensemble level, this effect is known as instantaneous spectral diffusion, but here at the single ion level, we will refer to it as external crosstalk. This effect can be limited by using optical pumping techniques to ensure that no spatially close ion is simultaneously close in frequency to the qubit resonance. The sequence to realize this is explained further in Sec. 4.1, when describing how to find more ions to scale the system up. Since the disturbing ion cannot have any of its levels close to any of the active levels of the qubit ion, it has to be further away than the spread of all hyperfine transitions. Simulations show that reserving a qubit bandwidth of $\pm 850$ MHz will ensure that the average error from external crosstalk is on the same order as the other errors listed. However, it is important to note that at the single ion level this error is random, and some qubits will not be affected at all while some qubits will be highly affected. In order to deal with the highly affected ions, one could use optical pumping schemes exploiting ion-ion interactions, in order to remove those ions (Wesenberg, Molmer, Rippe, & Kroll, 2007). It may also be possible to avoid this crosstalk by using non-random systems, such as stoichiometric crystals or systems with implanted ions, so further investigations into these techniques could be beneficial.

In addition to increasing the risk of crosstalk excitations, large pulse bandwidth can also contribute to using up more of the total spectral channels available to each qubit. If we allow different gate durations in the single- and two-qubit gate cases, then this does not restrict the single qubit gates much. Therefore, this mechanism is mainly discussed below for two-qubit gates and qubit nodes.

*Summary*

As described above, the fidelity of single qubit gates depends not only on materials properties but also on our choices of e.g. the desired number of qubits in a node. Trying to comply with all restrictions simultaneously, the limitations to the fidelity can be found considering the following:

1. Perform optical pumping to remove spectrally close ions, minimizing external crosstalk. The resulting spectral window will be determined by the level structure and will set an ultimate limit on the available gate bandwidth.
2. Investigate the number of qubits in a node as a function of pulse bandwidth: Check which pulse bandwidth can be used while maintaining a reasonable number of qubits and qubit connections, within the total addressable inhomogeneous bandwidth.
3. Check that the pulse bandwidth chosen in the step above gives an error due to internal crosstalk that is balanced with the error due to optical $T_2$ decoherence.
4. Optimize the pulse parameters within the allowed bandwidth to minimize decoherence from $T_2$, while using a pulse shape that does not introduce errors larger than those from $T_2$.

With these simultaneous restrictions, simulations of the Lindblad master equation performed on multi-level single ions, gives an expected single qubit gate fidelity error of $\sim 2 \cdot 10^{-4}$ (Lund University, 2021). This fidelity is well beyond the requirements for surfaces codes, and right at the threshold for full Steane-type error correction.

*Improving the fidelity*

It is important to note that the fidelity estimates and values listed here represent only a snapshot into a current activity. There is still a substantial potential for improvement of the fidelity of optimized protocols that may reach well into the regime where Steane-type error correction applies. Here, we list a few possibilities:

- Optimal control theory may identify excitation pulses that employ transition pathways that interfere and turn the internal crosstalk into a strength rather than an error source. This then may allow higher bandwidth gates and could further allow considering more types of RE ions as qubits, such as Pr which is currently limited in bandwidth by the internal crosstalk. Although the bandwidth might still be limited by the spectral pit structure which is also dependent on the levels.
- Take advantage of the multi-level nature of each RE ion to implement qudits with a larger Hilbert space. This would allow some multi-qubit gates like CNOT or Toffoli to be made fully internally, achieving fidelities closer to the single qubit gates. Bear in mind that one could use magnetic fields to split the levels of e.g. Eu to create such a scenario.
- Engineering RE systems, e.g. by using stoichiometric crystals together with point defects (Ahlefeldt, McAuslan, Longdell, Manson, & Sellars, 2013) or ion implantation (Groot-Berning, et al., 2019), could improve the situation substantially by ensuring that any spatially close ion is very far away in frequency. This would substantially reduce the error due to external crosstalk and could at the same time allow the use of larger bandwidths.
- Finding a better material with longer optical $T_2$, e.g. by using hosts consisting of atoms with no or very little spins, although this is not trivial since the material needs to fulfil many requirements as this section has discussed. $CaWO_4$ could be an example of such a crystal, if grown using isotopically enriched starting materials, or $TiO_2$ (Phenicie, o.a., 2019) using the magnetic interactions.

- Find a material where the ions have larger permanent dipole moments, since this would increase the ion-ion interaction shift, allowing the gates to use a larger bandwidth without limiting the number of qubits.
- Finally, we note that it is possible to take advantage of the specific physical interactions and employ the qubit connections beyond binary and nearest neighbor interaction in direct multi-qubit gates such as Toffoli gates. Such gates are heavily relied on, e.g., for some error correcting schemes, and their direct implementation and the optimization of their fidelity may be pursued with the same combination of physical reasoning and optimal control as we have applied to one- and two-qubit gates.

## 3.2 Two-qubit gates

### 3.2.1 Electric Dipole-dipole interaction

Rare earth ions have different permanent electric dipole moments in their ground and excited states, and the dipole-dipole interaction is strong enough to create entanglement between pairs of ions, subject to resonant excitation pulses (Ohlsson, Mohan, & Kröll, 2002). Many ideas can be shared between REIs and neutral alkali atoms which display similar dipolar interactions that can be switched on and off by excitation into highly excited Rydberg states (Mokhberi, Hennrich, & Schmidt-Kaler, 2020; Saffman, Walker, & Mølmer, 2010). We may for example employ a variant of the so-called Rydberg excitation blockade mechanism in which one ion, is resonantly excited from one of its ground state qubit levels. Depending on the qubit state, the conditionally excited dipole creates a surrounding electric field which shifts the excitation frequency of a neighboring ion as shown in Figure 6. Resonant driving of the second ion may thus cause or not cause transitions between its quantum states and the excited state amplitudes may be subsequently returned to the ground states resulting in a ground state qubit entangled state, or the process may be used to carry out a two-qubit gate.

The fidelity of the scheme depends on the ability of one ion to cause a large enough frequency shift to suppress excitation of the second ion. Thus, the ions must be spatially close enough that the shift exceeds the gate bandwidth (~1 MHz). For $^{153}$Eu:YSO with spatial separations of ~2-5 nm and doping concentrations of the order of a few percent, we expect a sizable number of ions to experience shifts of the order of 10-75 MHz. These are still typically smaller than the frequency separation (~100 MHz) of other internal levels of the ions. As will be discussed in more detail in Sec. 4.1, the cited parameters lead to the ability of one qubit to control the evolution of 5-10 other qubits.

We have described the blockade type gate, which relies on a detuning shift large enough to suppress excitation of the target ion. As an alternative scheme, we may employ the so-called interaction gate, which permits simultaneous excitation of a pair of ions. While they are excited, their mutual interaction causes a conditional phase to be accumulated on the doubly excited state. In principle, the states may be excited so fast that the blockade shift does not prevent the excitation, and in that case we obtain shorter gate times than required for the blockade entanglement scheme. The shorter time would then give a smaller fidelity loss due to optical decoherence. Unlike the blockade gate, such interaction gates rely on the precise accumulation of the phase shift and are hence not robust against variations in the interaction energy, but this may be a calibration issue and less of a problem with single instance quantum computing. The interaction gate protocol (assuming simultaneous population of the excited states) may also be improved by optimal control, and adiabatic versions may be derived from similar protocols for Rydberg atom quantum computing (Rao & Mølmer, 2014), and optimized by the shortcut-to-adiabaticity methods. Further inspiration may be derived from the Rydberg gate literature, such as the use of dark state combinations within the excited state manifold of states for two-qubit and multi-qubit gates (Petrosyan, Motzoi, Saffman, & Mølmer, 2017) (Khazali & Mølmer, 2020) which, however assume degeneracy of the atomic transitions and will thus need some adaption for application to dopant ions.

*Two-qubit gate fidelity*

Using the first CNOT gate protocol described above, a gate employing a conditional frequency shift in the optimal range ensures a minimum loss of fidelity. The CNOT gate can be constructed by a sequence of 4 pulses, 1) excite control ion, 2+3) a 2-pulse single qubit gate on the target ion (e.g. using the dark state scheme with cut-Gaussian pulses), 4) deexcite control ion. The single qubit gate in this sequence will be slightly more restricted on bandwidth compared to pure single qubit addressing, because the ion-ion shift needs to be larger than the bandwidth and using too high bandwidth per qubit will lead to fewer available qubits. Optimizing for a balance between fidelity and the number of qubit connections, we get a single qubit gate fidelity error to be $4 \cdot 10^{-4}$ in this case (Lund group, unpublished). From the sequence we see that the target ion spends one single qubit gate duration in the excited state and the control ion spends the equivalent of three gate durations in the excited state. Thus, if we assume that we are limited by optical $T_2$, then we should expect a combined CNOT gate error of four times the single qubit gate error, which would be $1.6 \cdot 10^{-3}$. This matches the simulation results quite well, and we generally believe that fidelity error should stay below about $2 \cdot 10^{-3}$ and remain robust over the full range of possible dipole shifts. Due to these relations, there is a trade-off between good two-qubit fidelity versus number of qubits and high connectivity. For more discussion concerning this see Section 4.1. Further details on the simulations are also discussed in the related SQUARE deliverable on the limiting mechanisms of high fidelity, D1.2.

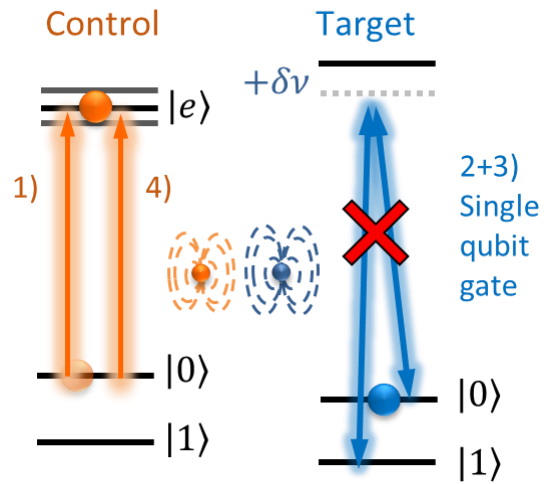

*Figure 6. Pulse sequence of the two-qubit gate. When the control is excited from the 0 state, the dipole blockade mechanism prevents the conditional gate on the target. See also the text for details.*

In general, many of the mechanisms that limit the two-qubit gate fidelity also limit the single qubit gate fidelity, and the improvements listed there will also improve the two-qubit gate.

### 3.2.2 Magnetic dipole-dipole interaction

Complementing the scheme based on electrical interactions as described above, magnetic interactions could also be used, allowing flexibility in our future choices of system. In many rare-earth materials, the magnetic dipole interaction can have a similar strength as the electric one. In the following we describe a system where this could be used instead, exploiting a central electron spin as means to entangle nearby nuclear spin qubits and to perform error corrections. This effect was demonstrated in a cluster of 3 nuclear spins weakly interacting with one NV center in diamond (Waldherr, o.a., 2014). Magnetic dipole-dipole interaction between the nuclei and the central spin was used to perform C-NOT gates with fidelities above 95%. This work shows that nearby nuclear spins can be used as a resource for quantum computation tasks.

A similar strategy can be used to perform two-qubit gates with REIs magnetically coupled to nearby nuclear spins. In fact, half of the $RE^{3+}$ ions have S=½ electron spin and, therefore, can play the same role as the NV center in the work mentioned above. This is depicted in Figure 7. Two-qubit gates can be performed between a) weakly interacting electron spins of two adjacent $RE^{3+}$ ions (same or different RE species, e.g. $Yb^{3+}$-$Yb^{3+}$ or $Yb^{3+}$-$Er^{3+}$) and b) the electron spin qubit and nearby nuclear spins, e.g. $Er^{3+}$-

$^{29}$Si in Y$_2$SiO$_5$. The central electron spin transition can be driven by resonant microwave radiation while single qubit gates on nuclear spins can be performed by RF pulses of certain area.

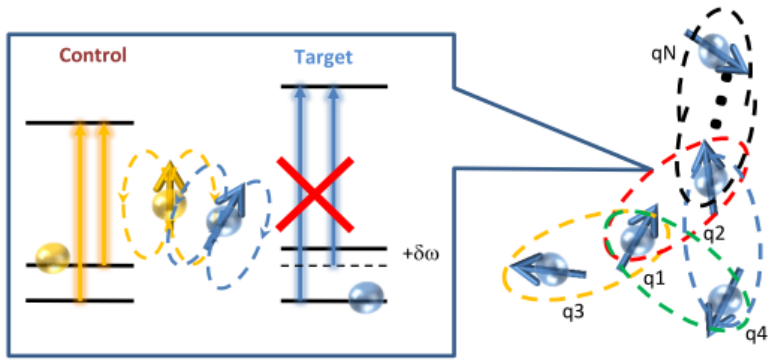

*Figure 7. A CNOT gate can be performed on a pair of closely spaced spin qubits communicating through the magnetic dipole-dipole interaction. A change in the spin state of the control qubit leads to the shift of spin transition of the target qubit.*

In case of interacting electron spins, two-qubit gates can be performed between the qubits having significantly different frequencies of their electron spin transitions similar to the strategy of entanglement generation in coupled pair of NV centers (Neumann, o.a., 2010). The spin state of the control qubit defines whether it is in resonance with the microwave π-pulse or not. For the species of the same kind, this can be achieved by picking the REIs in magnetically inequivalent sites such that in the external magnetic field they have distinct Zeeman splittings. Even in case of the same magnetic orientation, the two interacting REIs can be addressed separately if they have different optical frequencies. In this case, single-qubit gates can be performed on them independently, by exploiting a lambda-type optical driving similar to the protocols described for the gates based on the electric dipole moment described in the previous subsection. If the two interacting spins are of different RE species, they can be addressed independently by microwaves due to their distinct g-factors.

In order to eliminate the effect of microwaves onto other qubits besides the two on which the two qubit gate is performed, one performs the gate in the excited optical state whose g-tensor is significantly different from the one of the ground state. This is facilitated by long liftimes of REIs if they are not Purcell enhanced. The two qubits of interest are being excited with an optical π-pulse owing to spectral selectivity in the optical domain, their spins are then manipulated with microwaves, and later both are brought to the ground state with the second optical π-pulse. In a sense, one pulls two species of interest out of large ensemble, performs the gate only between them, and then brings them back to all other qubits. This situation is similar to dragging trapped ions one by one into interrogation region. This scalable method of performing gates on arbitrary qubit pairs requires, however, either rapid control of the Purcell effect, i.e. fast tuning of the cavity on and off resonance with the ions of interest, or enhancing only a dedicated readout ion.

The fidelity of the two-qubit gate is limited by both decay during the pulse duration and off-resonant interactions as a result of the interaction shift compared to the gate bandwidth, as described in more detail in Sec. 3.2.1. The magnetic dipole interaction is of similar strength as the electric one, so an interaction shift of the order of 10 MHz can obtained with similar separations as before, i.e. an electron spins separation of ~2-5nm. Thus, if the T$_2$ of the spin transition is around 1ms, a fidelity above 99.9% is achievable. Although, in a scenario with closely spaced spins, one must ensure that they do not cause uncontrollable shifts due to their magnetic field or dephasing due to spin flips, which could e.g. be done by controlling the separation by implantation as discussed in Sec. 6.2.3.

Similarly, C-NOT gates can be performed on nuclear qubits. The qubit state can be flipped by an RF π-pulse or not depending on the state of the control electron qubit. In this case, the hyperfine interaction between the nuclear and the central electron spin must be greater than the homogeneous linewidth of the nuclear spin transition.

### 3.2.3 Cavity-mediated interactions

As an alternative, or supplement, to the blockade and interaction gates among nearby ions, it is natural to consider schemes where interactions are mediated by the quantized radiation field of an optical cavity mode. Superconducting qubit architectures rely strongly on such interactions, but they also benefit from very large single-photon single-qubit interaction strengths ($g$). Proposals with trapped atoms and ions must carefully minimize the effects of atomic dissipation (at rate $\gamma$) and photon loss (at rate $\kappa$), and deterministic entanglement protocols using excitation transfer among the qubits and the field have an infidelity of order $\sqrt{(\kappa\gamma/g^2)}$. Lower infidelities can be achieved by schemes where the photon interacts dispersively with the qubits. In one such protocol for two ions where both are resonant with the cavity on the qubit level 1 – excited state transition, an incident photon will be reflected without entering the cavity if one or both ions occupy the qubit 1 level, while it will enter and experience the reflection as by an empty cavity if they both occupy the qubit 0 level. This scheme thus yields a phase gate (and detection of the photon may further increase the fidelity) (Wade, Mattioli, & Mølmer, 2016). The infidelities of these schemes, of order $\kappa\gamma/g^2$, are still insufficient using current generation of cavities for quantum computing with rare earth ions, and they need further development to deal with the different qubit frequencies, which might not both be resonant with the cavity mode. They do, however, provide promising prospects for scaling to systems with multiple interconnected cavities. Insufficient entanglement may be distilled by multiple rounds of transmission and local operations, and the interactions may also be enhanced by the use of ancillary qubits (see Sections 4 and 5).

## 4 Scaling to 100 qubits (single node NISQ processor)

This section elucidates on how the qubit-qubit mechanisms described in Sec. 3.2 can be used to form an (noisy) intermediate scale quantum processor node, known as a NISQ (Preskill, 2018). Currently, it seems technologically feasible that clusters of connected unique single ions would lead to a processor node of ~50-100 qubits, and it is described here how this can be achieved. The next section, Sec. 5, then discusses how the number of qubits can be extended beyond this range.

### 4.1 Electric Dipole-dipole interaction

The frequency shifts caused by the dipole-dipole interaction between ions used for the entangling gates, is also the mechanism used to find all ions that could be a part of a qubit node that makes up the intermediate quantum processor. Similarly to the high fidelity discussion, this section will also focus on $^{153}$Eu:YSO, if nothing else is specified. Not all Eu ions that gives large enough shifts can be used as qubits however, as some might be spectrally overlapping. Thus, any useful ions should fulfil the criteria:

1. Having a unique resonance frequency, such that they can be addressed in frequency space. This means that for a new ion to be added to the processor node as a useful qubit, neither its gate transitions, nor any of the other levels of the ions can overlap with any of the previous qubits' frequencies. The reason that the other levels also cannot overlap is because they would also excite from the ground state levels and increase the risk of external crosstalk on the qubits.

2. Being spatially sufficiently close that they can apply an interaction shift greater than the gate bandwidth. Here, one could use different bandwidths for single- and two-qubit gates, as described in Sec. 3.1.2, and in this case the shift must be larger than the two-qubit gate bandwidth, which for our $^{153}$Eu pulses are around 10 MHz. However, there is a tradeoff between the acceptable two-qubit gate error and the minimally allowed interaction shift.

### 4.1.1 Scheme to identify and isolate unique ions as qubits

The start of any search for qubit ions is to first identify a readout ion, and then find one single qubit ion that can control the readout. The first qubit ion is found by sequentially exciting frequency intervals of the qubit ions' inhomogeneous width until the fluorescence from the readout ion stops. After successively narrowing the interval to find the ion's exact resonance frequency, the first qubit with a sufficiently large shift has been found. After the first, there is more than one way to continue the search for the full node. Here we describe two such protocols that have slightly different outcomes. It is also important to note, that due to the high stability of the ion environment of these crystals, such a search procedure would only have to be done once for any given node.

*Line protocol*

Starting from the one ion that can control the readout, find *one* ion that can control the first qubit ion. This search is the same as for the first, i.e. scan the inhomogeneous width, except we now look for signs that the readout ion fluorescence is switched on again, revealing an ion that stopped the first ion from stopping the fluorescence. The line goes on by finding a third ion that can switch off the second, etc. Once no more ions (with unique spectral channels) can be found to control the latest qubit, one can go backwards and again search for more ions that can control the second to last etc, until all the remaining unused inhomogeneous width has been used up or no more controlling ions are found. This effectively creates a long line of qubits, with branches added late in the search process.

*Starfish protocol*

Starting from the one ion that can control the readout, the full inhomogeneous width is searched for *any* ions that can control this first ion before moving on. Then for each secondary ion found this way, the remaining inhomogeneous width is searched for any new ions that can control the secondaries etc. This effectively creates a starfish shape of the qubit connections, where most qubits sit close to each other.

Typical nodes of ions produced by the different protocols is visualized in Figure 8. In both protocols, the node is filled either when no more spectral channels are available, or when no new ions can be found that can control any of the previous ions. It should be noted that the entire inhomogeneous width does not need to be searched at every step. For each ion found, once its resonance frequency is known, the full spectral channel that it occupies is also known and can be ignored in future searches.

In both protocols, there may be other ions in the blocked frequency regions that cannot be used as qubits since they have some transition that overlaps with a qubit one. In order for these ions not to contribute to dephasing through external crosstalk, they should be moved to a ground state that lies outside of the spectral channel for any qubits, which can be done by optical pumping techniques.

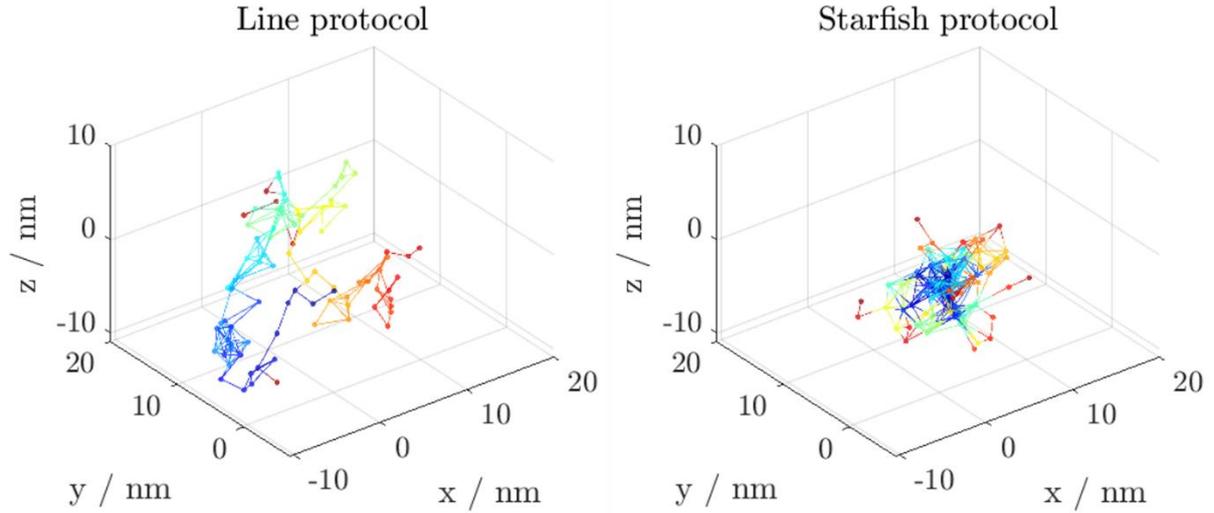

*Figure 8. Visualization of typical ion nodes for the Line and the Starfish search protocols, respectively, implemented as described in the main text. The figures are created by simulating the random distribution of dopants given a host crystal and their respective protocols (Lund University, 2021). The Starfish protocol produces on average more highly connected qubits, although the connections are unevenly distributed across the node. The color (ranging from blue to red) indicates when in the search the qubit has been found.*

### 4.1.2 Qubit connections and parameter dependence

From the level scheme of the qubit ions and the bandwidth of the gates, one can obtain the spectral channel that is blocked by each qubit. For the discussion, we designate the resonance frequency for each ion as its lowest energy transition, i.e. 1/2 ground state → 1/2 excited state. For $^{153}$Eu, using gate bandwidths that generate the high fidelity discussed in the section above, an interval from -850 MHz to +850 MHz compared to the resonance needs to be blocked. Thus, unique qubits could be spaced at best 850 MHz apart, and most likely about 1 GHz apart if the doping concentration is sufficient.

To achieve a large expected number of qubits in a single processor node, a large addressable inhomogeneous width is required. A large width can be created in the crystal be several different means, e.g. by using a high doping concentration, as it is known that the inhomogeneous width is approximately constant until about 0.1 % (For Eu:YSO) at which point the width increases about linearly with concentration. For a concentration of 5%, a width of about 100 GHz can be obtained (Könz, o.a., 2003). Other methods to increase the width includes co-doping with a defect species that distorts the lattice slightly, or by applying external fields with nm-sized electrodes to the crystal. In either case, the full width also needs to be addressable optically at a speed comparable to the gate speed, if the switching of the spectral channel should not be the main limitation of the processor. A technology to accomplish this has been investigated and is described in Sec. 7.1.

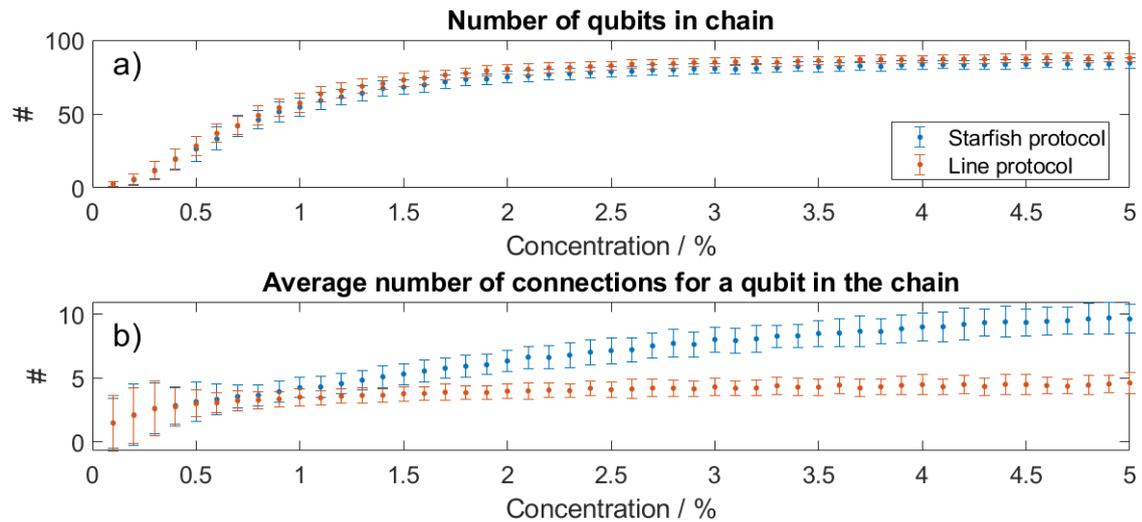

*Figure 9. Simulations of qubit connectivity for different search protocols (Lund University, 2021). a) Shows the number of total qubits obtained by the different protocols as a function of concentration. The spectral channel per qubit is roughly 1 GHz, and it is assumed that the laser can address qubits within a width of 100 GHz. b) Shows the average number of connections a qubit has to other qubits, also as a function of concentration. Here the differences between the two ion finding protocols become clearer. These simulations allow for two-qubit errors of up to 1e-2.*

The number of qubits in a node and their connectivity depend on how weak of a dipole-dipole interaction is allowed. This is in turn coupled to the two-qubit gate error, e.g. a dipole-dipole shift of at least 10 MHz results in the low two-qubit error as listed previously in Section 3.2.1 of ~2e-3, whereas a shift of about 2.5 MHz results in a higher two-qubit error of about 1e-2. There is thus a tradeoff of low two-qubit gate errors versus number of qubits and their connectivity.

In Figure 9 we show the concentration dependent number of qubits in a node where we accept possible errors of up to 1e-2 for some of the two-qubit connections. As can be seen in a) one can expect to reach up to about 100 qubits in a single node. It is already challenging to optically address 100 GHz with switching speeds comparable to the gate durations, but if addressing of an even larger width could be realized then the number of qubits per node would also increase correspondingly.

Both protocols have a similar number of total qubits for the given addressable width, and the protocols differ mostly in the distribution of qubit connections. For the Line protocol, there is on average a fewer number of qubit connections per qubit, although the distribution is quite even, meaning that each ion has a similar number of connections (75% of the qubits have around 4-6 connections). For the Starfish protocol, the closely spaced qubit selection gives on average a larger number of connections per qubit, especially at higher doping concentrations, although the connections are more unevenly distributed. This means that the first qubits (~5) you find will have many connections, >15, the following ~40 qubits will have about 10-15 connections, and the rest of the qubits will have about 2-10 connections.

If instead we would only accept the low two-qubit error of ~2e-3 and thus only allow dipole-dipole shifts larger than around 10 MHz, the total number of qubits found for a concentration of about 3-5% would drop to around 60-70, and the average number of connections per qubit would drop to around 3-4 (similar for both protocols, since the starfish protocol will almost become a line protocol due to fewer interactions per qubit on average). However, this drop might be mitigated by using two different types of two-qubit gate operations. For shifts larger than 10 MHz the dipole blockade interaction would still be used, but for small shifts one could consider using the interaction gates discussed in Section 3.2.1. This would have the benefit of allowing those low dipole-dipole interaction shifts when estimating the total number of qubits and their connectivity in a node while still maintaining a relatively low two-qubit gate error.

*Summary*
Overall, it seems feasible with current technology to achieve a node of about 100 qubits with an average of 10 qubit connections per qubit. Keep in mind that this number of qubits and connections depend on the acceptable two-qubit error. Some connections will have lower/higher two-qubit errors than others. This should allow a REQC to be a strong contender for a NISQ processor, where a variety of multi-qubit gate types could be implemented.

## 4.2 Magnetic dipole-dipole interaction

Similar considerations can be applied to magnetically dipolar coupled electron spins of REIs. As discussed in Section 3.2.2, single- and two-qubit gates can be performed on the electron spin qubits by optical means in a lambda-type configuration. In this scenario, individual ions can be addressed separately due to spectral selectivity in optical domain even though their spin transitions are very close in frequency space. Consequently, two-qubit gates can be performed on any coupled pair of REIs by picking the appropriate laser wavelengths within the inhomogeneous optical profile. In turn, coupled electron spins can form mutually coupled networks similar to the ones described in Section 4.1.2. In that sense, two-qubit gates mediated by Stark shifts induced by optical excitation of the control ion on the target one are complementary to the ones mediated by Zeeman shift induced by the control spin on the target spin. In both cases, multiplexing comes from an ability to optically segregate REIs within the inhomogeneous linewidth.

## 4.3 Cavity-mediated interactions

Unlike the line and starfish topologies that rely on spatially close ions, a cavity mode may in principle interact with all ions and mediate all-to-all entanglement and gate operations (see end of Section 3). The outstanding challenges are the reduced fidelity due to the single-photon single-ion coupling and the different frequencies of different ions precluding their simultaneous interaction with the cavity mode. One way to mitigate the different frequencies is to use modulation methods, where it is possible to equip a single cavity mode with sidebands to make it simultaneously and coherently interact with different frequency qubits (Andersen & Mølmer, 2015). Another way could be to use static local electric fields to shift the individual frequencies.

In the NISQ era, the gate fidelities may be sufficient for quantum simulation algorithms, and since entangled states with reduced fidelity can be purified, we imagine that the cavity may also be used for high fidelity entanglement, e.g., of members of different line or starfish clusters of ions. These may then permit distributed quantum processing over multiple sub-processors. On a final note, cavities mediate and vastly amplify the interactions between stationary qubits and light and may already in the short term play a crucial role for readout of qubit states. Use of ancillary ions may further improve these protocols both by the protection of the qubit information from fields at the cavity resonance frequencies (Zhong T. , Kindem, Rochman, & Faraon, 2017), by the choice of ions with stronger two-level coupling to the cavity field, and by the potential use of collective cavity coupling to ensembles of ancillary ions (Wade, Mattioli, & Mølmer, 2016) (Debnath, Kiilerich, & Mølmer, 2020).

# 5 Spin-photon interfaces and scaling to multi-node architectures

As described in the previous section, a NISQ processor node based on RE ions may achieve about 100 qubits coupled by mutual dipole-dipole interactions. In order to scale significantly beyond this number, other strategies that e.g. connect different nodes are required. In this section, we address the use of optical interfaces and communication by travelling qubit states of light between sub-processor units in multi-node architectures.

Interfaces between stationary and flying qubits are under investigation for all quantum computing candidate systems. Since ideas and analyses may be applied across different platforms, it may thus be useful to recall the recent progress in different systems and note the differences and similarities with the rare earth schemes.

It has been possible to demonstrate entanglement generation between super conducting qubits in different circuits connected by microwave waveguides (Dickel, o.a., 2018) and some of the protocols discussed in this report may be readily adapted to both circuit QED architectures and RE ions. While waveguides do not need cooling to the 10-100 mK regime for coherent transfer of quantum states (Vermersch, Guimond, Pichler, & Zoller, 2017), communication by microwaves is restricted to local networks (Magnard, o.a., 2020), and different protocols are being pursued to convert qubits with microwave excitation frequencies to optical photons, e.g., by the combination of hyperfine and optical transitions in rare-earth ion systems (Everts, Berrington, Ahlefeldt, & Longdell, 2019) or by the optical and microwave transitions into and among Rydberg states in atomic ensembles (Petrosyan, Mølmer, Fortágh, & Saffman, 2019).

Work on gate operations between photons and single trapped atoms and ions have progressed tremendously in recent years, see e.g. (Hacker, Welte, Rempe, & Ritter, 2016; Borne, Northup, Blatt, & Dayan, 2020). Quantum dots and other single photon sources in solid hosts, such as NV centers in diamond, are now routinely embedded in optical cavities and integrated optical waveguide structures for optimal interfacing and application with similar protocols (Lodahl, o.a., 2017), which may also be used in a rare-earth context.

Atomic ensemble memories with neutral atoms and ion doped crystals offer mechanisms to reliably capture and release single phase matched photons. The stored excitations do not interact in large ensembles, and quantum gate and entanglement operations are done in a heralded manner by detection of the released photons (Sangouard, Simon, de Riedmatten, & Gisin, 2011). Atomic ensembles of thousands of atoms may, however, be confined to 10 µm diameter ensembles across which the Rydberg blockade mechanism can effectively implement a strongly coupled "super atom" collective qubit degree of freedom. Such super atoms interact strongly with quantum light (Peyronel, o.a., 2012; Paris-Mandoki, o.a., 2017), and they are readily extended to study interactions and interferences of multiple storage states (Li, o.a., 2016).

Notably, by distillation of entanglement, imperfect quantum gates between matter and light do not prevent the establishment of high fidelity entangled states of remote qubits (Jiang, Taylor, Sørensen, & Lukin, 2007; Pedersen & Mølmer, 2009), and hence scalable quantum technologies, across larger network architectures (Nickerson, Fitzsimons, & Benjamin, 2014). The prospects to apply rare-earth ions in large optical multi-node architectures are thus excellent, and we shall discuss technical aspects as well as enabling methods for advanced operations of such architectures in the remainder of this Section.

The ability to apply optical cavities to achieve high-efficiency coupling between single rare-earth ions and single photons (see Sec. 2) opens prospect to connect single qubits in different nodes, forming a quantum network. Ultimately, separate interconnected nodes may be used to scale quantum computers to include thousands or millions of qubits in a distributed quantum computing architecture. The linking of more qubits and the stringent requirements on gate fidelities for quantum computing makes this a formidable long-term task for any quantum computing proposal. Several levels of complexity can be used for the network nodes, leading to different capabilities (Wehner, Elkouss, & Hanson, 2018). In this section, we describe how RE ions can be used as few qubit quantum processing nodes in a quantum network. It is important to note that for connecting nodes to increase the number of qubits, local networks are sufficient, e.g. the nodes could be in the same laboratory or even in the same cryostat. It is however gratifying that we may work towards this goal and at the same time address another prime application of quantum networks: distribution of entanglement to perform long distance quantum key distribution. For this application, it is in fact sufficient to consider heralded entanglement between remote qubits and local quantum gates between pairs of qubits, and the requirement on the gate fidelities are significantly lower than for quantum computing. Contrary to local networks, the distribution of photonics qubits over long distances in optical fibers require photons at telecom wavelengths.

## 5.1 Single RE ions quantum repeater nodes for long distance quantum communication.

A quantum repeater architecture allows the distribution of entanglement over very long distances. In a standard quantum repeater, the total distance is separated into several elementary links. Entanglement is first distributed in a heralded fashion within these elementary links and stored in quantum memories. Then, once two neighboring links are successfully entangled, a joint Bell State measurement is performed on the two qubits in the middle node, say B and C in the figure, which extends the entanglement to the remote qubits A and D. Successive entanglement swapping stages can then extend the distance further (see Figure 10).

In that context, rare-earth-ion-doped solids have been up to now mostly used as ensemble based quantum memories (Afzelius, Gisin, & de Riedmatten, 2015), where single photons are stored as collective atomic excitations (de Riedmatten, Afzelius, Staudt, Simon, & Gisin, 2008; Seri, o.a., 2017). The ability to control and manipulate single RE ions would, however, open new possibilities for developing quantum repeaters. The biggest advantage is that as we saw previously, high-fidelity quantum gates can be realized between single RE ion qubits. These two-qubit gates can be used to perform deterministic Bell state measurements, in contrast to ensemble-based approaches where the linear optic Bell state measurements are probabilistic, with an efficiency of 50%. This capability has an important impact on the achievable entanglement rate over large distances (>500 km), where repeaters with multiple swapping levels are required. Another important potential advantage of single RE ion qubits as quantum repeater nodes is that highly efficient spin-photon interfaces are possible, since RE ions can be embedded in nanoscale materials which allow the use of very small cavities with high Purcell enhancement. In addition, ions with different frequencies may be used to perform frequency multiplexed entanglement links, enabling a significant increase in rate. Finally, a great advantage is the possibility to achieve spin-photon interfaces at telecom wavelengths, using Er ions emitting at 1535 nm, in the telecom C band where the loss in optical fibers is minimal (0.2 dB/km).

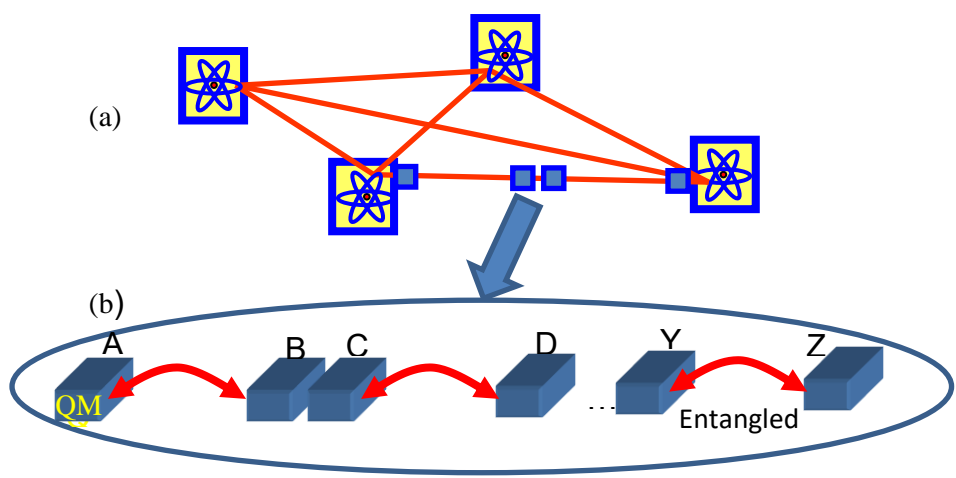

*Figure 10. (a) Example of a quantum network where small scale matter based quantum processors are interconnected optically using photons in optical fibers. (b) Quantum repeaters allowing the distribution of entangled states over distances much longer than the quantum channel attenuation. Entanglement is first distributed independently in each channel A-B, C-D, etc, and then stored in quantum memories (QM). When two adjacent segments are successfully entangled, a Bell State measurement is performed on the two middle QMs (e.g. B and C), which transfers the entanglement to the outer ensembles (e.g. A and D) and thus extends the entanglement distance. This scheme is then concatenated until A and Z are entangled.*

A specific quantum repeater protocol tailored for single rare-earth ions has been proposed in (Kimiaee Asadi, o.a., 2018). The protocol uses pairs of single ion qubits at each node, a communication qubit using a single Er ion, and a memory qubit using a single Eu ion. At each node, entanglement between Er ions and photons is created using the scheme introduced by (Barrett & Kok, 2005). Heralded entanglement between two remote Er ions within elementary links is performed by measuring the emitted photons. This entanglement is then transferred to the Eu ions for long-term storage and deterministic Bell state measurement using a two-qubit gate. It is important to note here that the requirement on the gate fidelities are significantly lower for quantum communication than for quantum computing. The threshold fidelity depends on the final fidelity desired for the remote qubits, on the number of swapping stages involved, and on the entanglement fidelity in the initial links $F_{link}$. For an $N$ link repeater, the final fidelity can be calculated as:

$$F = (F_{link})^N \times (F_{BSM})^{N-1}$$

A final remote two qubit entanglement fidelity of 78 % (with respect to a Bell state) is sufficient to violate a Bell inequality and to perform entanglement based QKD. For example, for a two link repeater, if we assume $F_{link} = 0.95$, a modest two qubit gate fidelity of 0.87 would be sufficient to distribute entanglement with a quality sufficient to violate a Bell inequality.

In order to generate entanglement between remote ions in an elementary link, the photons emitted by the ions must interfere at a beam splitter via the Hong-Ou-Mandel effect and must therefore be indistinguishable. This means they should have the same spectral and temporal profile and be coherent (Fourier transform limited). The last requirement is challenging with single rare-earth ions because of the long population relaxation lifetime of the excited state, and of the additional dephasing present, especially in nanostructures. The Fourier limited regime is reached when the Purcell enhancement C > $2\pi T_1 \gamma$, where $T_1$ and $\gamma$ are the excited state lifetime and the homogeneous linewidth of the emitter respectively. For realistic future cavity parameters, this regime should be readily achievable, as explained in Sec. 2.2. For early implementation one can note however, that it is not required to achieve fully coherent photon generation to perform heralded entanglement between two ions. By temporal gating, it is indeed possible to recover high-fidelity entanglement at the expense of a strong reduction in entanglement rate.

By multiple resend and distillation protocols, the entangled states may be useful, even in regimes where they are only produced (heralded) with very low probability, cf., studies with remote ion traps (Monroe & Kim, 2013) and with NV centers in diamond (Hensen, o.a., 2015).

## 5.2 Schemes for distributed quantum computing with rare earth ions

In Secs. 2, 3 and 4, we have presented components that permit, in principle, the construction of a network with a number of local quantum processors with up to ~100 connected qubits and a single or a few qubits reserved for communication with other similar processors. Optimization of algorithms for such a modular design should restrict the use of the comparatively slow communication operations to a minimum and explore the optimal number of qubits that is assigned for communication and short term storage of qubits entangled with qubits at other nodes. Since heralding by optical detection may play an important role for communication it seems natural to also integrate the measurements schemes for quantum error correction with the communication tasks in a platform independent theoretical study.

The coupling to cavities provides the direct interface of the qubit ions to the quantum internet. We have a choice between theoretical protocols, either where the carrier photons encode and transport qubit states, or where they serve as collective measurement degrees of freedom for heralded, non-local, entanglement operations on stationary qubits. We note that the co-doping of crystals with different ions may provide ideal memory qubits and communication qubits with distinct optical frequencies, and that the cavity systems also offer themselves as local processing and memory units (quantum repeaters) for long distance quantum communication.

While of a less universal nature, it is pertinent to employ also the specific advantages offered by any given physical implementation of qubits and light-matter interfaces. We shall thus observe and benefit from the specific short cuts to advanced operations offered by RE ions. As discussed in previous sections, the excited state interaction shifts apply between any number of control and target qubits within the dipolar interaction distance and hence a number of logical qubits can simultaneously affect a communication qubit and accomplish more complex operations. Also, the simultaneous interaction with several ions may permit a collective enhancement by multiple communication ions to couple a logical qubit to photons, and hence accomplish more complex network tasks in few operations. For some of these proposals to function (e.g., the one by (Wade, Mattioli, & Mølmer, 2016)), they should, however, be made insensitive to the inhomogenous frequency shifts of the ions, which could be accomplished e.g. by electric field tuning of small local electrodes. The largest gains in both near term and long term studies of optically interfaced rare-earth ions may come from development of the optical network and making it an integral part of both local and non-local gate protocols. Since the qubit state of a single ion in a cavity control the reflection and transmission properties for the cavity, it provides an interferometric element, a quantum switch, that controls the path and phase of the optical field. In (Wade, Mattioli, & Mølmer, 2016) it was demonstrated that two *non*-interacting ions in a cavity would undergo a controlled-PHASE gate upon reflection of a single photon by the cavity, and in the protocols described in (Cohen & Mølmer, 2018) scattering of a single-photon wavepacket on multiple cavities accomplishes multi-qubit C-PHASE and Toffoli gates on remote qubits in one single operation. Further work has been initiated to explore the use of separate ancilla and qubit ions for readout, communication, entanglement and quantum gate purposes (Debnath, Kiilerich, & Mølmer, 2020). This may be used directly in distributed quantum computing schemes, or they may be primarily reserved for dedicated purposes such as preparation of entangled states for teleportation of gates and states. Since these protocols are not universal but intimately connected with the states and interactions of the RE ions, their further theory development should go hand-in-hand with the experimental investigations.

### 5.3 Interfacing with microwave photons

Optical-to-microwave conversion has been proposed with the use of RE ions and other solid state dopants, embedded in optical and microwave cavities (Everts, Berrington, Ahlefeldt, & Longdell, 2019). These all rely on the strong collective coupling to many emitters, while coupling the qubit state of a single ion to a microwave photon may be more challenging. The blockade coupling of a single qubit ion to an ensemble of nearby ions may alter the conditional reflection of a microwave photon from a cavity, but the coupling may be orders of magnitude too weak. Recent work points to exponentially enhanced quantum interactions with parametrically driven cavity fields (Qin, et al., 2018) (Leroux, Govia, & Clerk, 2018), which may indeed be a path forward both for microwave and optical interfaces for single quanta.

## 6 Material considerations of rare-earth-ion systems

In the previous detailed sections, the techniques for reading out single ions, performing high fidelity gates, and connecting qubits are described, respectively. However, there are many combinations of RE ions, hosts, and material compositions, and it is a key challenge to find such combinations where all aspects are optimized simultaneously.

In particular, the main approach for reading out single ions relies on micro-structured cavities, while the primary requirement for high fidelity gates is a long $T_2$ together with high bandwidth. Thus, in this section, we focus on the aspects of finding a material system that can combine long $T_2$ with the nano-scale needed to reach the single ion regime.

## 6.1 Grown nanomaterials (bottom-up approaches)

RE doped nanoparticles and thin films can be synthesized by various methods. While bottom-up syntheses allow designing structures at the nm scale, with high flexibility in composition, shape etc., this approach also faces important challenges to preserve RE optical and spin coherence properties. Indeed, additional dephasing can occur in nanomaterials because of interactions with various defects, residual disorder, or size- and shape-related phonon modes.

### 6.1.1 Nanoparticles

Nanoparticles of $Y_2O_3$ obtained by homogeneous precipitation of an amorphous precursor in aqueous solution followed by high temperature annealing have shown optical and spin $T_2$ that are long for being nano-materials (Zhong & Goldner, 2019). So far, 5.7 µs optical $T_2$ ($\Gamma_h$ = 56 kHz) was observed in 100 nm diameter $Eu^{3+}$:$Y_2O_3$ particles at 1.4 K (Liu, o.a., 2020) and nuclear spin $T_2$ of 1.3 ms was measured in 400 nm particles of the same compound (Serrano, Karlsson, Fossati, Ferrier, & Goldner, 2018). In comparison, nanoparticles ground from bulk crystals show optical homogeneous linewidths in the MHz range (Utikal, o.a., 2014). Optical and spin values observed in bottom-up $Eu^{3+}$:$Y_2O_3$ are still about 100 and 10 times shorter than in bulk crystals, clearly evidencing additional decoherence processes. Similar results were reported in $Pr^{3+}$:$Y_2O_3$ nanoparticles (Serrano, o.a., 2019). An optical homogeneous linewidth of 380 kHz has been measured in $Er^{3+}$:$Y_2O_3$ nanoparticles of 150 nm diameter, at a temperature of 3 K and with a small magnetic field of 33 mT (unpublished). Further, in $Nd^{3+}$:$Y_2O_3$ nanoparticles a linewidth of 80 kHz was reached at 2 K and 300 mT. These spectroscopic studies revealed that the optical $T_2$ is limited first by Two-Level Systems (TLS), usually linked to residual disorder, and second by electric field fluctuations. Magnetic perturbations, which do not seem to play a role for optical transitions, except possibly in the case of $Er^{3+}$ that has large g factors, are otherwise identified as a limiting process for the spin transitions. Importantly, for particles as small as 100 nm, temperature dependent measurements did not reveal modified phonon modes, which could cause strong dephasing and set a hard limit on the size of useful systems. Several strategies to improve nanoparticles properties are currently considered. First, material synthesis and post-processing can be optimized to reduce defects such as oxygen vacancies and residual disorder. For example, an oxygen plasma treatment significantly improves optical $T_2$ in particles by modifying vacancy related defects (Liu, o.a., 2020). Other synthesis methods for which crystallization is directly obtained could potentially lead to higher crystalline quality by avoiding the need to remove organic groups during annealing. Lower annealing temperatures could also offer better perspectives for reducing dephasing induced by surface charges or dangling bonds by growing shells around particles. Other host crystals can be obtained as nanoparticles, such as $Y_2SiO_5$ or $YVO_4$, and could result in materials with reduced disorder and defects and thus longer $T_2$, at the expense of e.g. more difficult synthesis ($Y_2SiO_5$), a higher spin density and site symmetry not allowing permanent electric dipole moments ($YVO_4$). Second, specific experimental conditions could be used to reduce RE coupling to perturbations. As in bulk materials, clock transitions can reduce magnetic field dephasing. In the case of $^{171}$Yb, this occurs at zero magnetic field for optical and spin transitions (Ortu, o.a., 2018). In site symmetries where permanent dipole moments vanish, coupling to electric field noise is also expected to be largely reduced, which could be useful in magnetic interaction-based gates. Effects of TLS are reduced by decreasing temperature, with a contribution to $\Gamma_h$ that typically varies linearly with T. Increase of $T_2$ at mK temperatures has been reported in several systems (Staudt, o.a., 2006), and could be beneficial too for nanoparticles, although it is not clear to what extent current high-temperature values can be extrapolated.

### 6.1.2 Thin films

Grown thin films are another approach to nanomaterials that may allow long coherence lifetimes. They have several advantages, including a high control on the growth process in terms of thickness, purity and other parameters like doping profile. RE thin films for application to quantum technologies ($Y_2O_3$ and related materials) have been produced by Chemical Vapor Deposition (CVD), Atomic Layer Deposition (ALD), Pulsed Laser Deposition (PLD) and Molecular Beam Epitaxy (MBE) (Zhong & Goldner, 2019). Optical homogeneous linewidths have only been reported for CVD films, with values in the 10s of MHz range (Harada, o.a., 2020). This is attributed to a large amount of structural defects and indicates the need for higher crystalline quality. This could be obtained in epitaxial films grown on a suitable substrate such as silicon. By MBE, narrow $Er^{3+}$ optical inhomogeneous linewidths (down to 5 GHz) have been observed, although no homogeneous linewidths have yet been reported (Singh, o.a., 2020). The use of other more flexible growth methods, such as CVD, using appropriate conditions and specific substrates could constitute an advantageous alternative.

## 6.2 Nanomaterials obtained from bulk crystals (top-down approaches)

### 6.2.1 Milled crystals and properties of surface ions

RE ions in nano-resonators directly milled in single crystals have shown excellent properties, both for optical and spin transitions. As an example, 94 µs optical $T_2$ has been measured in Nd:$Y_2SiO_5$ photonic crystal cavities, which is essentially the bulk value (Zhong, o.a., 2015). Here, RE ions are at least about 75 nm away from the surface of the cavity structures. Another impressive result is the 30 ms spin $T_2$ in the same type of cavity using $^{171}$Yb:$YVO_4$ (Kindem, o.a., 2020).

Surface ions in bulk crystals can also show good properties, as shown in Pr:$Y_2SiO_5$. Evanescent coupling to a surface waveguide enabled probing ions within 130 nm from surface by photon echoes (Marzban, Bartholomew, Madden, Vu, & Sellars, 2015). Homogeneous linewidths down to a few kHz were measured in this way. Much broader linewidths, in the MHz range, were however measured for single $Er^{3+}$ ions coupled to cavities deposited on $Y_2SiO_5$ (Dibos, Raha, Phenicie, & Thompson, 2018). The origin of this broadening has not been established yet but could be due to electric noise induced by surface charges or dangling bonds. It could be mitigated using hosts in which RE permanent electric dipole moments vanish (e.g. $YVO_4$), however also preventing using the electric dipole gate mechanism. Alternatively, post-treatments could also modify surface termination and reduce electric noise.

Diffusion of RE ions into a undoped bulk crystal can be obtained from a doped layer deposited on the surface under high temperature annealing. This can result in ions located within about 100 nm from the surface that exhibit spectral holes with a few MHz width, presumably due to a high RE concentration (Ferrier, o.a., 2020).

### 6.2.2 Thin films from bulk crystals

A top-down approach to produce thin single crystalline films of RE-doped materials is to start with a single crystal slab of material, e.g. YSO, and polish it down to an appropriate thickness of 1-2 µm. Obviously, such polishing should be performed on the crystal bonded to an appropriate substrate which, in case of FP cavities, is a mirror. Alternatively, polishing on a dummy substrate and transferring on a mirror later can be attempted and has recently been successfully demonstrated down to a thickness of about 5 – 10 µm (Merkel, Ulanowski, & Reiserer, 2020). The strategies towards this bonding will be discussed in Sec. 7.3. Here we will discuss only the possible issues associated with polishing.

Coarse grinding of the crystal followed by chemical-mechanical planarization (CMP) is the process to be implemented to obtain thin films of RE-doped materials. Provided the right slurry is chosen to planarize the crystal surface, the roughness well below 1 nm on tens of µm scale can be achieved.

In analogy with NV centers in diamond, polishing introduces large number of defects into the material. Depending on the polishing quality, the defects can propagate from 10 µm up to 200 µm under the surface of the crystal. The defects probably include dislocations, line defects, etc. Aiming for the best coherence properties of the RE dopants in the film, the defective layer should be removed. Therefore, polishing needs to be stopped several microns (around 10 µm) before and then the top layer has to be removed by other means. The best way to do so is to dry etch the polished surface of the crystal with e.g. an argon plasma to remove the top several microns. Argon plasma is chosen to introduce only mechanical etching component and having no chemical part since later can propagate though the bulk of the crystal and deteriorate the film quality even further. Considering typical rates of argon milling of RE oxides (10-100 nm/min), removing the defective layer can be quite a lengthy procedure. However, the quality of the obtained film is expected to be the best in terms of coherence lifetime and in terms of surface flatness since argon milling tends to planarize the surface further.

### 6.2.3 Ion implantation

Doping of the material with REIs can be done during crystal growth, as discussed above, or *a posteriori* by implanting them in desired locations under the crystal surface. The latter route has the advantage of high spatial localization of REIs with a well-controlled concentration (Groot-Berning, o.a., 2019). In that work, it was shown that it is possible to implant individual REIs extracted from a Paul trap and focus them onto the surface of a crystalline host with an accuracy of 50 nm, and with a production yield of 50%. It is also possible to implant ions through lithographically defined hard masks. The implantation depth is defined by the ion energy and can vary between a few nanometers for keV and a few microns for MeV energy ranges.

On the other hand, ion implantation produces damage in the crystalline structure that affects coherence properties of REIs, which needs to be removed by high temperature annealing. The latter is not always possible, especially in case of thin implanted films bonded to a substrate. The limiting annealing temperature is typically defined by the difference in thermal expansion coefficients of the substrate and the film. For example, implantation of lithium niobate with REIs requires annealing at 1060°C to completely remove implantation-induced defects (Fleuster, Buchal, Snoeks, & Polman, 1994). However, thin films on Si/SiO$_2$ substrate cannot be annealed at temperatures above 650°C. Treatment at higher temperatures leads to cracks on the surface of the film.

Having this in mind, the procedure of creating well-defined patterns of RE implants can be viewed as follows. First, one side of a clean slab of the crystal is polished, planarized, and the polishing damage is removed by argon milling (see Sec. 6.2.2). After that, the polished surface is implanted with the desired pattern by either masked implantation or by focused ion beam. This is followed by high temperature annealing of the slab to remove implantation damage. This is later followed by a bottom-up approach of creating the mirror and the substrate directly on the surface of the slab (see Sec. 7.3 below). Finally, the other surface is thinned down, polished, planarized, and argon milled. The result of this quite lengthy route is a thin (sub-micron) film of a single crystalline material with well-defined patterns of RE implants placed on highly reflective dielectric mirror supported by an electroplated metal holder.

### 6.3 Concerning readout ion – qubit ion pairs

In order to ensure that the cavity enhancement does not introduce any fast decay on the qubit excited states that are used for qubit-qubit interactions, the primary way of scaling a REQC is to separate the task of reading out and acting as qubit to different RE species. This also gives the benefit that two ions can be chosen that optimizes their respective task. For instance, Eu currently appears to have the best coherence properties and a level structure that allows reasonable gate bandwidths, as discussed above. Since only very few readout ions are needed, it would be possible to use only the trace amounts of those readout ions, in a fully doped crystal with Eu. Several different readout ions would be possible with reasonable high readout fidelities as described in Sec. 2. For instance, both Nd and Er, which is at the telecom wavelength, would give high emission rates. One challenge with both of these is that they have an electron spin. This spin produces a local magnetic field which could split the levels of qubit ions that are sufficiently close, which is known as super hyperfine interaction. Any splitting of more than ~1 kHz could impact the fidelity of the gates, by reducing the available bandwidth, although there are several ways to mitigate this. One way would be to use so far unexplored gates based on numerical optimization, such as optimal control, in order to include the split levels in the gates, thus removing the bandwidth limitations. In a similar vein, the split levels could also be turned into an advantage, if it would be possible to use the extra levels to encode a higher Hilbert space dimension. Further, it could be possible to regain the unsplit level structure by applying a stronger external magnetic field. Using a field of the few Tesla, the Eu ion would split a few 10s of MHz, which is of the same order of as the separation between the nuclear qubit states. This means that a similar gate bandwidth could be used to again obtain a reasonably high gate fidelity. Finally, a readout ion without an electron spin can also be chosen, such as Pr or Tm, although the expected photon emission rate would probably drop slightly, in the order of a factor ~5 (for Pr).

Another concern is if there are any spectral overlaps between the readout and qubit ions. This could lead to energy transfer mechanisms, and if the qubit ion got excited by such a process it could lose its coherence. However, energy transfer mechanisms has been investigated (Serrano, et al., 2014) and due to the $1/R^6$ scaling, it mostly affects ions at very close ranges, less than ~1 nm of each other. And even if one qubit ion happens to be so close to the readout, one could simply ignore that ion as a qubit, removing it by optical pumping. Spectral overlap on the excitation wavelengths could also be a problem, e.g. spectra from (Altner, Wild, & Mitsunaga, 1995) suggests a possible overlap between the main line of Eu at 580 nm and excitation from the ground state of Nd to one of its higher crystal field levels. Although, in the $Y_2SiO_5$ host there are two usable sites, and typical wavelength differences between the sites are of the order of nm, so most likely the other site would work even if there happened to be an overlap for the first. For Er or Pr, no such overlap is expected, although a full experimental investigation of any cross-species transfers would be valuable to confirm the available pairings for quantum information purposes.

## 6.4 Summary of material approaches

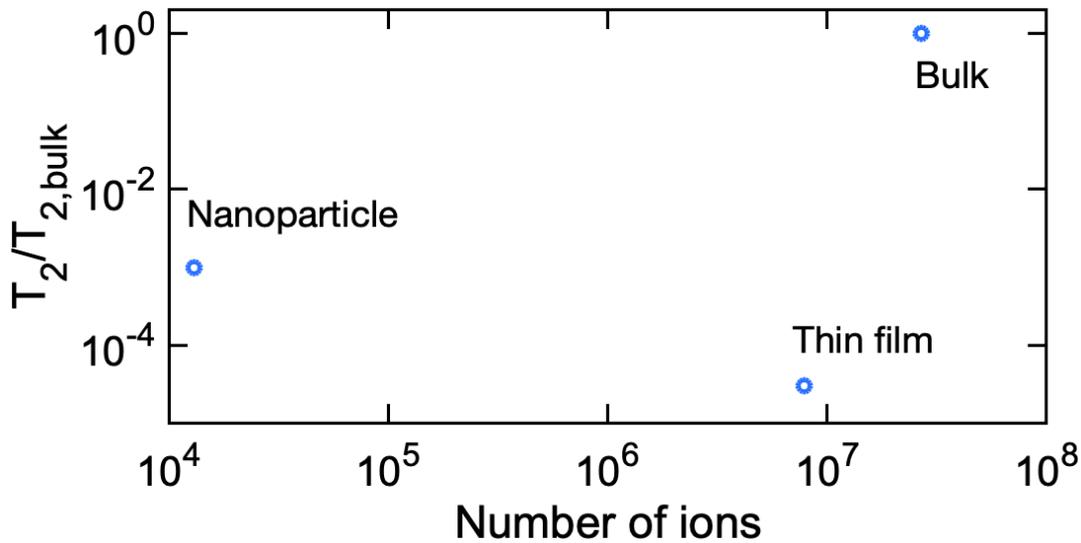

*Figure 11. Number of ions in a 1 µm³ volume and optical coherence lifetimes for rare earth doped crystals under different forms. A homogeneous doping is assumed, at a level of 1 atomic percent. The nanoparticle and thin film have a diameter and thickness of 50 nm respectively, and the bulk crystal (refractive index n = 1.8) fills the cavity volume. $T_2$ values are estimated from measurements on $Eu^{3+}:Y_2O_3$ nanoparticles and thin films (Zhong & Goldner, 2019).*

Figure 11 shows a comparison between different material approaches in the case of a fiber based optical cavity. All materials are assumed homogeneously doped at a 1% doping level, giving an average distance between ions of about 1.7 nm in $Y_2O_3$ which allows a high two-qubit interaction strength. The required inhomogeneous broadening, 100 GHz, could be reached by further co-doping the materials at a level of a few at. % with optically 'inert' ions such as $Sc^{3+}$ which have no electron spins and therefore limited impact on coherence lifetimes (Welinski, o.a., 2017). The vertical axis shows optical coherence lifetimes relative to bulk crystal, which is important for gate fidelity. Nanoparticles offer the smallest volume and therefore combine a high number of interacting qubits while preserving single qubit spectral selectivity. A bulk crystal thinned down to 1 µm would contain about $10^3$ more ions than a nanoparticle, thus providing more potential qubits but also making addressing single qubits difficult. The spectral resolution of the qubit ions needs to be around 1 kHz to allow high fidelity gates. With an inhomogeneous linewidth of 100 GHz, and $10^7$ ions in the focal spot, there is then about 0.1 ion per kHz. Thus, it should be possible to separate the qubit ions from background ions, by means of optical pumping, although the pumping pulses need be long, in the order to ms. It can also be noted that the optical pumping could be performed quickly by making use to stimulated emission to crystal field levels, followed by spontaneous phonon decay to the groundstate (Lauritzen, Hastings-Simon, de Riedmatten, Afzelius, & Gisin, 2008).

The benefit of a bulk crystal is a much longer optical $T_2$ than nanoparticles. Indeed, even if nanoparticles are so far the best nanomaterials in terms of $T_2$, they suffer from significant additional dephasing processes ($T_2/T_{2,bulk} \approx 10^{-3}$). The strategies outlined above (see Sec. 6.1.1) could reduce these processes and make nanoparticles a suitable platform for a medium size RE quantum computer. Grown thin films are a promising alternative (see Sec. 6.1.2) but should be optimized to increase coherence lifetimes from the current values ($T_2/T_{2,bulk} \approx 10^{-5}$). One advantage is that they can be grown with localized doping in nm thick layers, which can reduce the number of ions in the cavity volume, while ensuring strong qubit interactions. This approach of localized doping can also be applied to bulk crystals using ion implantation or surface diffusion (see Sec. 6.2.3). Demonstration of long coherence lifetimes has, however, not yet been reported in these cases.

# 7 Enabling technologies

The future success of quantum computing platforms in general, and indeed for RE platforms in particular, will depend on the availability of certain technologies, such as optics, laser sources or cryogenics. In this section we go through some of those technologies that are of key importance to the RE systems and propose concrete short-term technological solutions to overcome the current limitations.

## 7.1 Laser system to address many qubits

### 7.1.1 Requirements of such a system[3]

As mentioned before, intermediate scaling using qubit-qubit mechanisms requires optically control of individual rare earth ions, which are distinguishable by their frequency. Having the possibility to address a larger bandwidth will lead to a larger number of potential qubits. One estimates that addressing a 10 GHz bandwidth (resp. 100 GHz) will allow the realization of 10 (resp. 100) qubits. The main challenges for the laser source are then:

- the capability to emit short (typ. close to the gate speed ~1 µs), arbitrary-shaped (optimal control), at a high repetition rate (closely spaced) over the largest possible bandwidth,
- the requirements on signal to noise ratio (SNR) and spurious extinction (high fidelity of the operations),
- the wavelength of operation: the europium ion being the preferred qubit candidate, the laser source should emit at 580 nm.

### 7.1.2 Idea and feasibility of building the laser system in practice

The current experiments used dedicated lasers designed for single qubit gates, with acousto-optical modulators (AOM). This architecture suffers from lack of scalability, versatility and performances in order to achieve multi-qubits operation:

- limited tuning range and speed,
- limited bandwidth for waveform,
- limited flexibility, each laser source being dedicated for the targeted ion in a specific experiment, i.e. at a specific wavelength.

Due to the requirements on arbitrarily shaped, closely spaced pulses being separated by arbitrary frequency steps, a solution using a single tunable laser source is very challenging. Indeed, to the best of our knowledge, there has been no demonstration of a laser source exhibiting such a high tuning speed (100 GHz/µs). Moreover, if existing, such a tuning speed would need to be compatible with a fast stabilization, in order to ensure the requirement on phase and frequency stability/repeatability from pulse to pulse. The laser would then be inserted in a phase-locked-loop (PLL) and be also limited by the performances of electronics. The highest tuning range for a high performance PLL being below 1 GHz/µs, it is orders of magnitude lower than the requirements for quantum computing using RE ions.

---

[3] In this paragraph, we take the example of an Eu-based system, but the architecture works with most of the commonly used REI (especially for $Pr^{3+}$ and $Er^{3+}$) by changing the up-converted wavelength.

An architecture based on the combination of multiple externally modulated, fixed-frequency laser sources can be an efficient solution. At the telecommunication wavelength of 1.5 µm, it is possible to obtain the required performances in terms of arbitrary pulse shaping, SNR and spurious suppression by using commercially available components (high speed arbitrary waveform generators, $LiNbO_3$ IQ optical modulators, narrow optical filters and wavelength multiplexers) over a multi-GHz bandwidth. By combining ~3 lasers, addressing 10 GHz is possible. The conversion of the signal at Europium wavelength is achievable by sum frequency generation in efficient PPLN wavelength up-converters, which are also commercially available with a technologically realistic pump laser source at ~1 µm being required. It is also crucial that the output signal at 580 nm exhibits a high frequency stability. Several schemes can be envisaged by using highly stable reference optical frequency combs, on which either the 1.5 µm laser sources, the pump laser source and/or their SFG combination are frequency locked.

To reach 100 GHz bandwidth, the combination of several tens of lasers will be necessary. Even if 30 lasers can be integrated in a 6U 19" rack, which is quite common for telecom or sensing applications, the scalability could benefit further from the recent developments of photonic integrated chips (PIC). In that case, it is also very likely that only one pump laser would not be enough to efficiently frequency up convert all the lasers. Tens of channels could be implemented in a small footprint package, but losses and spurious suppression issues are not yet demonstrated at the level of discrete components. We can expect that in the future, the performances (linewidth, spurious, SNR, filtering roll-off) will improve and be comparable to that of discrete components. Particularly, the development of $LiNbO_3$ on insulator PIC platform would be extremely attractive for this application, and other quantum technologies fields, since it would offer both highly linear phase modulation (giving high quality waveforms) and on-chip wavelength conversion.

## 7.2  Cryogenics to reach low temperatures and high mechanical stability

Existing closed cycle cryogenic platforms currently used for quantum experiments in the 2-4 K range are based on commercial cryo-coolers (i.e. He cooling machines around which are built cryostats) most of them being from non-EC origin (SHI Cryogenics, Jp, Cryomech, US). The only EC provider of cooling machines at 4 K is TransMIT (GE), which proposes a 2-stage PT machine with attractive performances, but exhibiting high vibration levels requiring complex damping architectures for e.g. cavity-based experiments. Moreover, the He compressors driving the above-mentioned machines are not compatible with potential embedded applications. Finally, the minimum temperature is usually limited to 2.7 K, which is not sufficient for some quantum platforms.

The characteristics of existing solutions are then not satisfying in terms of performances versus efficiency, size and vibration level, which can be critical for the future scalable products and potential embedded systems (e.g. space-based).

An analysis of the constraints required for quantum computing using rare-earth ions in micro-cavities has then been conducted to propose and explore in details several possible architectures and their pros and cons, while keeping in mind European-based development, scalable product and future out-of-lab environment. They are described in details in the Deliverable D3.7 of the SQUARE project of the Quantum Flagship. They are mainly composed of usual stages: Pulse Tube (PT), Gifford Mac-Mahon (GM), He3 or He4 Joule-Thomson (JT) stages. The best compromise for a compact system and possibly out-of-laboratory on a medium-time scale is a He4 JT 3K stage combined with a continuous Adiabatic Demagnetization Refrigerator (cADR) or a He3 JT. The achieved cooling power (~10 mW at 1K) could be sufficient in most cases, by optimizing the optical heat load at the sample area through the use of reflective materials and escape windows. This system can share the He4 JT stage of the 3 K architecture, and very low temperature range is accessible (<1.5 K, needed for Kramers ions such as Er3+). For a complete system, it is also extremely critical that the proposed architectures should be combined with a low vibrations damping stage in order to meet the requirements in terms of vibrations, as already mentioned in this roadmap.

Consequently, the main developments to foreseen in Europe is to combine a 4He Joule-Thomson with a low vibration stage, and to demonstrate a compact system combining 4He Joule-Thomson and a continuous Adiabatic Demagnetization Refrigerator (cADR).

Current developments for harsher environment are already on-going in Europe and are compatible with the proposed architectures of this deliverable and then with quantum computing using rare-earth ions. The efforts should be pursued in order to get closer to a product on a medium time scale.

This analysis focuses on REI requirements, but other technologies could benefit from these cryogenic developments, for example, quantum light sources based on 2D materials, quantum networks with NV and SiV centers in diamond and ensemble quantum memories in rare earths.

## 7.3 Bonding thin films onto mirrors

Growing directly thin films on mirrors is only possible at low deposition temperatures (below 400 ºC) to avoid mirror damage, which in turn may result in poor film properties such as an amorphous structure or very broad optical transitions.

If one exploits a top-down approach of creating thin films discussed in Sec. 6.2.2, two different bonding techniques can be used.

The first one is the direct bonding of the planarized (see Sec. 6.2.2) but still thick crystal to the existing mirror. After bonding, the opposite (unbonded) surface of the crystal has to be polished down to the appropriate thickness and planarized with CPM and argon milling as the last step. However, the ability to polish the bonded crystal depends on the bonding strength. In turn, the latter strongly depends on the planarization quality on the millimeter scale, appropriate activation of the surfaces of both the mirror and the crystal, cleanness of the environment, etc. These considerations make this technique quite challenging.

Alternatively, one can use bottom-up approach to create the mirror and the substrate on the planarized surface of the crystal. The mirror can be sputtered directly on the surface of the crystal providing excellent bonding strength since now it is defined by covalent or ionic bonds rather than Van-der-Waals bonds exploited in the former case. The macroscopic roughness of the crystal surface will not play any role in the adhesion. After the deposition of the mirror, the substrate can be created also using the bottom-up route by depositing a seed layer of metal, e.g. copper, and electroplating it until the metal layer is hundreds of microns thick. The residual stress in the electroplated copper can be removed by mild annealing in vacuum at 200-300°C.

As an alternative to the previously described approaches that make use of a thinned bulk single crystal, one could use a layer (ideally epitaxial) that is directly grown on an appropriate substrate by various methods such as MBE, PLD or CVD for example. This would alleviate the need for precision polishing and allow a greater flexibility in material composition and dopant spatial localization. For telecom wavelengths, MBE-grown crystalline mirror coatings from GaAs/AlGaAs may allow for combined growth of mirror coatings and REI crystals. Nevertheless, the high deposition temperatures required by those methods to achieve reasonably good crystalline quality (in general T>500°C) precludes the direct growth on standard high reflectivity Bragg mirrors (composed of amorphous e.g. $SiO_2/TiO_2$ multilayers).

The films could thus be lifted off and, thanks to an adapted wafer carrier (such as PMMA), stamped onto the mirror cavity. Lifting off can be obtained by fully etching the substrate using wet or dry chemistries. KOH solutions or SF6 plasma are commonly used for isotropic and anisotropic etching of silicon for example. Another common approach is to rely on a sacrificial layer that is intercalated between the wafer and the rare-earth oxide film and that will facilitate its quick release by preferential etching. Inspiration from recipes developed in microelectronics for Silicon-On-Insulator (SOI) can guide such technological efforts.

Another great asset of thin grown films is the ability to use them in wave-guiding architectures by engineering stacked layers that allow light confinement. Optical cavities or resonators could be directly deposited onto the film in a similar way as demonstrated for bulk crystals, thus enabling high localization of RE ions with respect to the photonic structures.

## 7.4 Cavities, fabrication, photonic integration

The different cavity types described above rely on very different fabrication processes, control needs, material compatibility, and system integration.

Fiber cavities can be produced in a fast (<1min / fiber) and automatized manner by $CO_2$ laser machining, including predictive quality control, and the applied mirrors achieve state of the art quality guaranteed by highly optimized commercial IBS coating technology. A large range of materials including nanocrystals and bottom-up or polished down thin films can be introduced. Fully fiber-based cavities can be realized with minimal control overhead and passive stability (Saavedra, al., & Meschede, 2020), which have good prospects for scalability and remote operation.

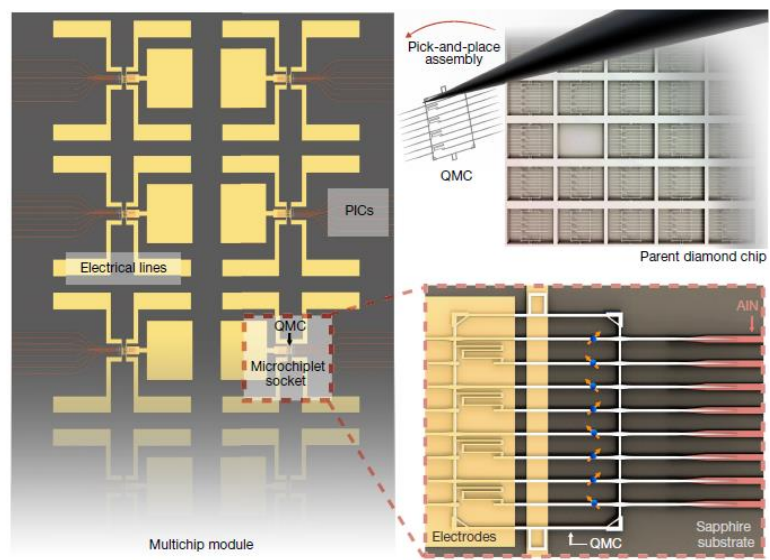

*Figure 12. A schematic drawing of a scalable hybrid photonic integrated circuit based on the separate fabrication of subcomponents with pick-and-place assembly. Taken from (Wan, al., & Englund, 2020).*

FIB-milled nanocavities involve only a single fabrication step which is however somewhat more time consuming, but large numbers of cavities could be produced on a single crystal. Any bulk crystal or thin film can be used, and in principle, all wavelengths are accessible, even though visible wavelengths demand highest quality structures. Outcoupling is typically done via free-space collection off-chip, which requires cryogenic nanopositioning and thus complicates the setup.

Nanophotonic cavities produced by e-beam lithography and reactive ion etching can be produced in large numbers on silicon-on-insulator wafers. Transfer to the REI crystal is non-trivial but can possibly be automated. Outcoupling can be done non-contact into single-mode fibers, which however requires cryogenic nanopositioning.

In the long term, the largest potential lies in the full integration on larger-scale photonic chips, possibly including the laser source, modulators, waveguides coupled to tunable cavities with incorporated RE ions, dynamic signal routing, and efficient waveguide-coupled superconducting single photon detectors. The rapid progress in this field has recently led to remarkable results with color centers in diamond integrated into a large-scale hybrid photonic circuit as seen in Figure 12 (Wan, al., & Englund, 2020). There, a 128-channel, defect-free array of germanium-vacancy and silicon-vacancy colour centers in diamond waveguides was integrated in an aluminium nitride photonic integrated circuit. The emitters showed Fourier-limited emission lines, efficient coupling to waveguides, and wide frequency tuning of emission frequencies by local strain. This example illustrates the fabrication perfection and integration flexibility of today's state-of-the art devices, and when extrapolated to future RE devices with $10^4$ individual cavities hosting $10^2$ qubit ions each, one may imagine a 1-million-qubit REI processor.

*Acknowledgements*

This Roadmap was written as a part of the SQUARE project, as a part of the EU Flagship on quantum technologies (grant agreement No 820391). We gratefully acknowledge comments from Mikael Afzelius and Ferdinand Schmidt-Kaler. P.B., L. M. and S.W. acknowledges fruitful discussions with Thierry Trollier and Philippe Camus from Absolut System.